\documentclass[12pt]{article}
\usepackage{amssymb,amsmath,epsfig}
\usepackage{graphicx}
\usepackage{amsfonts,graphicx,amssymb,amsmath,amsthm,epsfig}
\usepackage{float}
\allowdisplaybreaks

\begin{document}
\title{\bf Generalized Ghost Dark Energy Model in Extended Modified Theory of Symmetric
Telleparallel Gravity}

\author{M. Sharif \thanks {msharif.math@pu.edu.pk}~ and Iqra Ibrar
\thanks{iqraibrar26@gmail.com}\\
Department of Mathematics and Statistics, The University of Lahore,\\
1-KM Defence Road Lahore-54000, Pakistan.}
\date{}
\maketitle
\begin{abstract}
This paper focuses on developing a generalized ghost dark energy
model $f\mathcal{(Q,T)}$ through the correspondence scheme for a
non-interacting scenario involving pressureless matter and a
power-law scale factor. We analyze the cosmological implications of
our model using the equation of state parameter,
$\omega_{GGDE}-\omega'_{GGDE}$ and the $r-s$ planes. We also examine
stability of the model by considering squared speed of sound
parameter. The equation of state parameter shows the phantom era,
while the squared speed of sound indicates a stable generalized
ghost dark energy model throughout the cosmic evolution. The
$\omega_{GGDE}-\omega'_{GGDE}$ plane depicts a freezing region,
whereas the $r-s$ plane corresponds to the Chaplygin gas model. We
conclude that our results are consistent with the current
observational data.
\end{abstract}

\section{Introduction}

In the early 20th century, Albert Einstein introduced the general
theory of relativity (GR), which revolutionized our comprehension of
the cosmos. It is expanding through a significant accumulation of
accurate observations and delving into the concealed realms of the
universe in current cosmology. The current concept of cosmic
expansion has been reinforced by several observable findings,
including large-scale structures, cosmic microwave background
radiations and supernovae type Ia \cite{1}. The universe in which we
live is experiencing a unique epoch, in which it expands at an
accelerated rate. Cosmologists examined this expansion by conducting
numerous attempts on distant galaxies. It is assumed that this
universe expansion is due to an extraordinary force called dark
energy (DE), which works through immense negative pressure. Several
models have been proposed to explain phenomena related to DE and the
evolution of the universe. There are two main approaches to studying
DE, i.e., dynamical DE models and modified theories of gravity
\cite{2}.

In recent years, scholars have proposed various approaches to
address these issues, yet they remain enigmatic to this day. To
characterize DE, researchers found the equation of state (EoS)
parameter beneficial for the model being examined. The Veneziano
ghost DE (GDE) model possesses noteworthy physical properties within
the cosmos \cite{3}. In flat spacetime, the GDE does not contribute
to vacuum energy density but in curved spacetime, it enhances small
vacuum energy density in proportion to $\Lambda_{QCD}^3 H$, where
$\Lambda_{QCD}$ represents the quantum chromodynamics (QCD) mass
scale and $H$ denotes the Hubble parameter. With $\Lambda_{QCD}\sim
100 MeV$ and $H\sim 10^{-33}eV$, $\Lambda^{3}_{QCD}H$ gives the
order of observed GDE density. This small value efficiently provides
the essential exotic force driving the accelerating universe,
thereby resolving the fine-tuning issue. \cite{4}. The energy
density of GDE is given as
\begin{equation*}\label{39}
\rho_{GDE}=\alpha H,
\end{equation*}
where $\alpha$ is an arbitrary constant.

The GDE model solves various issues effectively, but it encounters
stability problems \cite{5}. It suggests that the energy density
relies not only on $H$ explicitly but also on higher-order terms of
$H$, leading to what is known as the generalized GDE (GGDE) model.
In this model, the vacuum energy associated with the ghost field can
be regarded as a dynamic cosmological constant. In reference
\cite{6}, the author explained that the Veneziano QCD ghost field's
contribution to vacuum energy does not precisely follow the
$H$-order. The additional term $H^{2}$ is particularly important
during the initial phase of the universe evolution, serving as early
DE \cite{8}. Instead, a subleading term $H^{2}$ arises because the
conservation of the energy-momentum tensor holds independently
\cite{7-a}. The GGDE density is given as follows
\begin{equation}\label{a}
\rho_{GGDE}=\alpha H+\beta H^{2},
\end{equation}
where $\beta$ is another arbitrary constant.

Numerous DE models can be explored to study the ongoing accelerated
expansion of the universe. Khodam et al. \cite{9-a} investigated the
reconstruction of $f\mathcal{(R,T)}$ gravity (here $\mathcal{R}$ and
$\mathcal{T}$ are the Ricci scalar and trace of the energy-momentum
tensor, respectively) within the GGDE model, delving into cosmic
evolution through the analysis of cosmological parameters. Malekjani
\cite{11-a} explored the cosmographic aspects of GGDE and concluded
that this model demonstrates strong consistency with observational
data. Ebrahimi et al. \cite{10} studied the interacting GGDE within
a non-flat cosmos, revealing that the EoS parameter aligns with the
phantom era of the universe. Chattopadhyay \cite{13} reconstructed
the QCD ghost $f(T)$ model ($T$ represents the torsion scalar) to
examine the cosmic evolution and found both the phantom and
quintessence phases of the cosmos. Sharif and Saba \cite{61-a}
explored the cosmography of GGDE in $f\mathcal{(G, T)}$
($\mathcal{G}$ is the Gauss-Bonnet invariant) gravity. Biswas et al.
\cite{61-b} investigated the evolution trajectories of the EoS
parameter, $\omega_{D}-\omega'_{D}$ and the state finder planes in
GGDE DGP model. Recently, Sharif and Ajmal \cite{61} studied the
cosmography of GGDE $f(\mathcal{Q})$ theory.

Several gravitational theories, including GR, are acquiring
importance because of their ability to explain the accelerating
expansion of the universe. The Levi-Civita connection in GR explains
gravitational interactions in Riemannian space based on the
assumption of geometry free from torsion and non-metricity.
Moreover, it is important to consider that the general affine
connection has a broader formulation \cite{12b}, and GR can be
derived within alternative spacetime geometries beyond Riemannian.
Teleparallel gravity presents an alternative framework to GR, where
gravitational interaction is characterized by torsion, denoted as
$T$ \cite{12c}. Teleparallel equivalent of GR utilizes the
Weitzenb$\ddot{o}$ck connection, which entails zero curvature and
non-metricity \cite{12d}. In the study of a cosmological model
within Weyl-Carten (WC) spacetime, the Weitzenb$\ddot{o}$ck
connection is investigated with vanishing of the combined curvature
and scalar torsion \cite{12e}.

In a torsion-free space, gravity is modified by non-metricity,
defined as
$\mathcal{Q}_{\lambda\zeta\xi}=\nabla_{\lambda}g_{\zeta\xi}$. This
theory is called symmetric teleparallel gravity (STG) \cite{12g}.
Jimenez et al. \cite{41} recently introduced $f(\mathcal{Q})$ theory
also named as non-metric gravity, the scalar function $\mathcal{Q}$
represents the non-metricity. Lazkoz et al. \cite{7} described the
constraints on $f(\mathcal{Q})$ gravity by using polynomial
functions of the red-shift. The energy conditions for two distinct
$f(\mathcal{Q})$ gravity models have also been discussed \cite{31}.
The behavior of cosmic parameters in the same context was examined
by Koussour et al. \cite{9}. Chanda and Paul \cite{11} studied the
formation of primordial black holes in this theory.

Xu et al. \cite{20} extended $f\mathcal{(Q)}$ gravity to
$f\mathcal{(Q,T)}$ gravity by adding trace of the energy-momentum
tensor into $f\mathcal{(Q)}$ theory. The Lagrangian governing the
gravitational field is considered to be a gravitational function of
both $\mathcal{Q}$ and $\mathcal{T}$. The field equations of the
corresponding theory are derived by varying the gravitational action
with respect to both the metric and the connection. The goal of
introducing this theory is to analyze its theoretical implications,
its consistency with real-world experimental evidence, and its
applicability to cosmological scenarios. N$\acute{a}$jera and
Fajardo \cite{18} studied cosmic perturbation theory in
$f\mathcal{(Q,T)}$ gravity and found that non-minimal coupling
between matter and curvature perturbations have a major influence on
the universe evolution. Pati et al. \cite{46-a} investigated the
dynamical features and cosmic evolution in the corresponding
gravity. Mandal et al. \cite{18a} studied the cosmography of
holographic DE $f\mathcal{(Q,T)}$ gravity.

In this paper, we use a correspondence technique to recreate the
GGDE $f\mathcal{(Q,T)}$ model in a non-interacting scenario. We
investigate cosmic evolution using the EoS parameter and phase
planes. The paper is structured as follows. In section \textbf{2},
we introduce $f\mathcal{(Q,T)}$ gravity and its corresponding field
equations. In section \textbf{3}, we explore the reconstruction
procedure to formulate the GGDE $f\mathcal{(Q,T)}$ paradigm. In
section \textbf{4}, we analyze cosmic behavior using the EoS
parameter and phase planes. We also examine the stability of the
resulting model. Finally, we discuss our findings in the last
section.

\section{The $f\mathcal{(Q,T)}$ Theory}

This section presents the basic structure of modified
$f(\mathcal{Q,T})$ theory. In this theory, the framework of
spacetime is the torsion-free teleprallel geometry, i.e.,
$R^{\rho}_{~\eta\gamma\nu}=0$ and $T^{\rho}_{\gamma\nu}=0$. The
connection in the WC geometry is expressed as \cite{20}
\begin{equation}\label{8}
{\hat\Gamma^{\lambda}_{\alpha\beta}}=\Gamma^{\lambda}_{\alpha\beta}
+\mathbb{C}^{\lambda}_{\alpha\beta}+\mathbb{L}^{\lambda}_{\alpha\beta},
\end{equation}
where $\Gamma^{\zeta}_{~\gamma\alpha}$ is the Levi-Civita
connection, $\mathbb{C}^{\lambda}_{\alpha\beta}$ is the contortion
tensor and $\mathbb{L}^{\lambda}_{\alpha\beta}$ is the deformation
tensor, given by
\begin{align}\label{5}
\Gamma^{\zeta}_{\mu\alpha}&=\frac{1}{2}g^{\zeta\sigma}
(g_{\sigma\alpha,\mu}+g_{\sigma\mu,\alpha}-g_{\sigma\mu,\alpha}),
\end{align}
\begin{equation}\label{11}
\mathbb{C}^{\lambda}_{\alpha\beta}=\hat\Gamma^{\lambda}_{[\alpha\beta]}
+g^{\lambda\phi}g_{\alpha\kappa}\hat\Gamma^{\kappa}_{[\beta\phi]}
+g^{\lambda\phi}g_{\beta\kappa}\hat\Gamma^{\kappa}_{[\alpha\phi]},
\end{equation}
and
\begin{equation}\label{12}
\mathbb{L}^{\lambda}_{\alpha\beta}=\frac{1}{2}g^{\lambda\phi}(\mathcal{Q}_{\beta\alpha\phi}
+\mathcal{Q}_{\alpha\beta\phi}-\mathcal{Q}_{\lambda\alpha\beta}),
\end{equation}
where
\begin{equation}\label{13}
\mathcal{Q}_{\lambda
\alpha\beta}=\nabla_{\lambda}g_{\alpha\beta}=-g_{\alpha\beta},_{\lambda}
+g_{\beta\phi}\hat{\Gamma}^{\phi}_{\alpha\lambda}
+g_{\phi\alpha}\hat{\Gamma}^{\phi}_{\beta\lambda}.
\end{equation}
The gravitational action can be written as \cite{30a}
\begin{equation}\label{20}
S=\frac{1}{2k} \int g^{\alpha\beta}(\Gamma^{\rho}_{\phi\alpha}
\Gamma^{\phi}_{\beta\rho} -\Gamma^{\rho}_{\phi\rho}
\Gamma^{\phi}_{\alpha\beta})\sqrt{-g}d^{4}x.
\end{equation}
Using the anti-symmetric relation
$(\Gamma^{\mu}_{\varsigma\alpha}=-\mathbb{L}^{\mu}_{~\varsigma\alpha})$
in Eq.$(\ref{20})$, the integral action takes the form
\begin{equation}\label{21}
S=-\frac{1}{2k} \int g^{\alpha\beta}(\mathbb{L}^{\rho}_{\phi\alpha}
\mathbb{L}^{\phi}_{\beta\rho} - \mathbb{L}^{\rho}_{\phi\rho}
\mathbb{L}^{\phi}_{\alpha\beta}) \sqrt{-g} d^{4}x.
\end{equation}
This action is known as the action of STG which is equivalent to the
Einstein-Hilbert action. There are some significant differences
between two gravitational paradigms. One of them is that the
vanishing of curvature tensor in STG causes the system to appear as
flat structure throughout. Furthermore, the gravitational effects
occur due to variations in the length of vector itself, rather than
rotation of an angle formed by two vectors in parallel transport.

Now, we look at an extension of STG and take an arbitrary function
of $\mathcal{Q}$, the above action takes the form
\begin{equation}\label{22}
S=\int\bigg[\frac{1}{2k}f(\mathcal{Q})+L_{m}\bigg]\sqrt{-g}d^{4}x,
\end{equation}
where
\begin{equation}\label{23}
\mathcal{Q}=-g^{\alpha\beta}(\mathbb{L}^{\rho}_{\mu\alpha}\mathbb{L}^{\mu}_{\beta\rho}
-\mathbb{L}^{\rho}_{\mu\rho}\mathbb{L}^{\mu}_{\alpha\beta}),
\end{equation}
which leads to $f(\mathcal{Q})$ theory. If we couple this theory
with the trace of the energy-momentum tensor, we can obtain
$f(\mathcal{Q,T})$ theory whose action is given as
\begin{equation}\label{19}
\mathcal{S}=\frac{1}{2k}\int f(\mathcal{Q,T}) \sqrt{-g}d^{4}x+ \int
L_{m}\sqrt{-g}d^{4}x,
\end{equation}
here $L_{m}$ represents the matter lagrangian. The traces of
non-metricity tensor are given by
\begin{align}\label{25}
\mathcal{Q}_{\rho}= \mathcal{Q}^{~\alpha}_{\rho~\alpha}, \quad
\tilde{\mathcal{Q}}_{\rho}= \mathcal{Q}^{\alpha}_{\rho\alpha}.
\end{align}
The superpotential in view of $\mathcal{Q}$ is written as
\begin{align}\label{26}
P^{\rho}_{\alpha\beta}&=-\frac{1}{2}\mathcal{L}^{\rho}_{\alpha\beta}
+\frac{1}{4}(\mathcal{Q}^{\rho}-\tilde{\mathcal{Q}}^{\rho})
g_{\alpha\beta}- \frac{1}{4} \delta ^{\rho}_{~[\alpha
\mathcal{Q}_{\beta}]}.
\end{align}
Furthermore, the expression of $\mathcal{Q}$ obtained using the
superpotential becomes
\begin{align}\label{27}
\mathcal{Q}=-\mathcal{Q}_{\rho\alpha\beta}P^{\rho\alpha\beta}
=-\frac{1}{4}(-\mathcal{Q}^{\rho\beta\zeta}\mathcal{Q}_{\rho\beta\zeta}
+2\mathcal{Q}^{\rho\beta\zeta}\mathcal{Q}_{\zeta\rho\beta}
-2\mathcal{Q}^{\zeta}\tilde{\mathcal{Q}}_{\zeta}+\mathcal{Q}^{\zeta}\mathcal{Q}_{\zeta}).
\end{align}

Taking the variation of $S$ with respect to the metric tensor as
zero yields the field equations
\begin{align}\nonumber
\delta S&=0=\int \frac{1}{2}\delta
[f(\mathcal{Q,T})\sqrt{-g}]+\delta[L_{m}\sqrt{-g}]d^{4}x \\\nonumber
0&=\int \frac{1}{2}\bigg( \frac{-1}{2} f g_{\alpha\beta} \sqrt{-g}
\delta g^{\alpha\beta} + f_{\mathcal{Q}} \sqrt{-g} \delta
\mathcal{Q} + f_{\mathcal{T}} \sqrt{-g} \delta
\mathcal{T}\bigg)\\\label{28}&-\frac{1}{2} \mathcal{T}_{\alpha\beta}
\sqrt{-g} \delta g^{\alpha\beta}d^ {4}x.
\end{align}
Furthermore, we define
\begin{align}\label{29}
\mathcal{T}_{\alpha\beta} &= \frac{-2}{\sqrt{-g}} \frac{\delta
(\sqrt{-g} L_{m})}{\delta g^{\alpha\beta}}, \quad
\Theta_{\alpha\beta}= g^{\rho\mu} \frac{\delta
\mathcal{T}_{\rho\mu}}{\delta g^{\alpha\beta}},
\end{align}
which implies that $ \delta \mathcal{T}=\delta
(\mathcal{T}_{\alpha\beta}g^{\alpha\beta})=(\mathcal{T}_{\alpha\beta}+
\Theta_{\alpha\beta})\delta g^{\alpha\beta}$. Inserting the
aforementioned factors in Eq.\eqref{28}, we have
\begin{eqnarray}\nonumber
\delta S &=&0=\int \frac{1}{2}\bigg\{\frac{-1}{2}f
g_{\alpha\beta}\sqrt{-g} \delta g^{\alpha\beta} +
f_{\mathcal{T}}(\mathcal{T}_{\alpha\beta}+
\Theta_{\alpha\beta})\sqrt{-g} \delta g^{\alpha\beta}
\\\nonumber
&-&f_{\mathcal{Q}} \sqrt{-g}(P_{\alpha\rho\mu}
\mathcal{Q}_{\beta}~^{\rho\mu}- 2\mathcal{Q}^{\rho\beta}~_{\alpha}
P_{\rho\mu\beta}) \delta g^{\alpha\beta}+2f_{\mathcal{Q}} \sqrt{-g}
P_{\rho\alpha\beta} \nabla^{\rho} \delta g^{\alpha\beta}
\\\label{30}
&+&2f_{\mathcal{Q}}\sqrt{-g}P_{\rho\alpha\beta} \nabla^{\rho} \delta
g^{\alpha\beta} \bigg\}- \frac{1}{2}
\mathcal{T}_{\alpha\beta}\sqrt{-g} \delta g^{\alpha\beta}d^ {4}x.
\end{eqnarray}
Integration of the term $ 2 f_{\mathcal{Q}} \sqrt{-g}
P_{\rho\alpha\beta}\nabla^{\rho}\delta g^{\alpha\beta}$ along with
the boundary conditions takes the form $ -2 \nabla^{\rho}
(f_{\mathcal{Q}} \sqrt{-g} P_{\rho\alpha\beta})\delta
g^{\alpha\beta}$. The terms $f_{\mathcal{Q}}$ and $f_{\mathcal{T}}$
represent partial derivatives with respect to $\mathcal{Q}$ and
$\mathcal{T}$, respectively. Finally, we obtain the field equations
as
\begin{align}\nonumber
T_{\alpha\beta}&=\frac{-2}{\sqrt{-g}} \nabla_{\rho}
(f_{\mathcal{Q}}\sqrt{-g} P^{\rho}_{\alpha\beta})- \frac{1}{2} f
g_{\alpha\beta} + f_{\mathcal{T}} (\mathcal{T}_{\alpha\beta} +
\Theta_{\alpha\beta}) -f_{\mathcal{Q}} (P_{\alpha\rho\mu}
\mathcal{Q}_{\beta}~^{\rho\mu}
\\\label{31} &-2\mathcal{Q}^{\rho\mu}~_{\alpha} P_{\rho\mu\beta}).
\end{align}
Its covariant derivative yields the non-conservation equation as
\begin{align}\nonumber
\nabla_{\alpha}T^{\alpha}_{~\beta}&=\frac{1}{f_{\mathcal{T}}-1}\bigg[\nabla_{\alpha}
\bigg(\frac{1}{\sqrt{-g}}\nabla_{\rho}H^{~\rho\alpha}_{\beta}\bigg)-\nabla_{\alpha}
(f_{\mathcal{T}} \Theta^{\alpha}_{~\beta})
-\frac{1}{\sqrt{-g}}\nabla_{\rho}
\nabla_{\alpha}H^{~\rho\alpha}_{\beta}\\\label{32}
&-2\nabla_{\alpha}A^{\alpha}_{~\beta}
+\frac{1}{2}f_{\mathcal{T}}\partial_{\beta}\mathcal{T}\bigg],
\end{align}
where hyper-momentum tensor density is defined as
\begin{equation}\label{81-a}
H^{~\rho\alpha}_{\beta}=\frac{\sqrt{-g}}{2}f_{\mathcal{T}}\frac{\delta
\mathcal{T}}{\delta
\hat{\Gamma}^{\beta}~_{\rho\alpha}}+\frac{\delta\sqrt{-g}L_{m}}
{\delta\hat{\Gamma}^{\beta}~_{\rho\alpha}}.
\end{equation}

\section{Reconstruction of GGDE $f\mathcal{(Q,T)}$ Model}

In this section, we use a correspondence technique to recreate the
GGDE $f\mathcal{(Q,T)}$ model. The line element of the isotropic and
spatially homogeneous universe model is given by
\begin{equation}\label{33}
ds^{2} = -dt^{2}+ \mathrm{a}^{2}(t)[dx^{2}+ dy^{2}+dz^{2}],
\end{equation}
where the scale factor is represented by $\mathrm{a}(t)$. The
configuration of isotropic matter with four-velocity fluid
$u_{\mu}$, pressure ($P_{M}$), and normal matter density
($\rho_{M}$) is given as
\begin{equation}\label{33a}
\tilde{\mathcal{T}}_{\mu\nu}=(\rho_{M}+P_{M})u_{\mu}u_{\nu}+P_{M}g_{\mu\nu}.
\end{equation}
The modified Friedmann equations in the context of
$f\mathcal{(Q,T)}$ gravity are expressed as
\begin{align}\label{34}
3H^2 &=\rho_{eff}=\rho_{M}+\rho_{GGDE},
\\\label{34-a}
2\dot{H}+3H^2&=P_{eff}=P_{M}+P_{GGDE},
\end{align}
where $H =\frac{\dot{\mathrm{a}}}{\mathrm{a}}$. The dot demonstrates
derivative with respect to cosmic time $t$. The non-metricity
$\mathcal{Q}$ in terms of the Hubble parameter is
\begin{align}\label{80}
\mathcal{Q}=-\frac{1}{4}\big[-\mathcal{Q}_{\rho\xi\eta}\mathcal{Q}^{\rho\xi\eta}
+2\mathcal{Q}_{\rho\xi\eta}\mathcal{Q}^{\xi\rho\eta}-2\tilde{\mathcal{Q}}^{\rho}
\mathcal{Q}_{\rho}+\mathcal{Q}^{\rho}\mathcal{Q}_{\rho}\big].
\end{align}
Simplifying this equation leads to
\begin{equation}\label{86}
\mathcal{Q}=6H^{2}.
\end{equation}
Furthermore, $\rho_{GGDE}$ and $P_{GGDE}$ denote the density and
pressure of DE, respectively, given as
\begin{align}\label{35}
\rho_{GGDE}&=\frac{1}{2}f\mathcal{(Q,T)}-\mathcal{Q}f_{\mathcal{Q}}
-f_{\mathcal{T}}(\rho_{M}+P_{M}),\\\label{36}
P_{GGDE}&=-\frac{1}{2}f\mathcal{(Q,T)}+2f_{\mathcal{QQ}}H+2f_{\mathcal{Q}}\dot{H}
+\mathcal{Q}f_{\mathcal{Q}}.
\end{align}
The conservation equation \eqref{32} takes the following form for an
ideal fluid
\begin{equation}\label{37}
\dot{\rho}_{M}+3H(\rho_{M}+P_{M})=\frac{1}{(f_{\mathcal{T}}-1)}
\bigg[2\nabla_{\beta}(P_{M}\mu^{\beta}f_{\mathcal{T}})
+f_{\mathcal{T}}\nabla_{\beta}\mu^{\beta}\mathcal{T}
+2\mu^{\beta}\mathcal{T}_{\alpha\beta}\nabla^{\alpha}
f_{\mathcal{T}\mu^{\beta}}\bigg].
\end{equation}
The first field equation \eqref{34} leads to
\begin{equation}\label{38}
\Omega_{M}+\Omega_{GGDE}=1,
\end{equation}
where $\Omega_{M}=\frac{\rho_{M}}{3H^2}$ and
$\Omega_{GGDE}=\frac{\rho_{GGDE}}{3H^2}$ represent the fractional
energy densities of normal matter and dark source, respectively.
Dynamic DE models with energy density proportional to Hubble
parameter are crucial to explaining the accelerated expansion of the
universe.

Next, we will use a correspondence approach in an ideal fluid
configuration to create the GGDE $f\mathcal{(Q,T)}$ model with
emphasis on the dust case $(P_{M}=0)$.

\subsection{Non-interacting GGDE $f\mathcal{(Q,T)}$ Model}

Here, we consider the standard $f\mathcal{(Q,T)}$ function in the
following form \cite{29}
\begin{equation}\label{40}
f\mathcal{(Q,T)}=f_{1}\mathcal{(Q)}+f_{2}\mathcal{(T)},
\end{equation}
where $f_{1}$ and $f_{2}$ depend upon $\mathcal{Q}$ and
$\mathcal{T}$, respectively. In this scenario, it is evident that
there is a minimal coupling between curvature and matter
constituents. This version of the generic function reveals that the
interaction is purely gravitational and hence easy to handle. This
can effectively explain the ongoing expansion of the universe.
Moreover, the reconstruction methodology demonstrates that such
generated models are physically viable \cite{29}. Using dust fluid
and Eq.\eqref{40}, the field equations \eqref{34} and \eqref{34-a}
yield
\begin{equation}\label{41}
3H^{2}=\rho_{eff}=\rho_{M}+\rho_{GGDE}, \quad
2\dot{H}+3H^{2}=P_{eff}=P_{GGDE},
\end{equation}
where
\begin{align}\label{42}
\rho_{GGDE}&=\frac{1}{2}f_{1}\mathcal{(Q)}+\frac{1}{2}f_{2}\mathcal{(T)}-\mathcal{Q}
f_{1\mathcal{Q}}+ f_{2\mathcal{T}}\rho_{M}, \\\label{43}
P_{GGDE}&=-\frac{1}{2}f_{1}\mathcal{(Q)}-\frac{1}{2}f_{2}\mathcal{(T)}
+2f_{1\mathcal{Q}}\dot{H}+2
f_{1\mathcal{QQ}}H+\mathcal{Q}f_{1\mathcal{Q}}.
\end{align}
The associated conservation equation \eqref{37} reduces to
\begin{align}
\dot{\rho}_{M}+3H\rho_{M}=\frac{1}{f_{2\mathcal{T}}-1}\big[2\mathcal{T}
f_{2\mathcal{TT}}+f_{2\mathcal{T}}\dot{\mathcal{T}}\big].
\end{align}
This equation is consistent with the standard continuity equation
when the right-hand side is assumed to be zero, implying
\begin{equation}\label{45}
{\dot\rho_{M}}+3H\rho_{M}=0\quad\Longrightarrow\quad
\rho_{M}=\rho_{0}\mathrm{a}(t)^{-3},
\end{equation}
with the constraint
\begin{equation}\label{46}
f_{2\mathcal{T}}+2\mathcal{T}f_{2\mathcal{TT}}= 0,
\end{equation}
whose solution provides
\begin{equation}\label{47}
f_{2}(\mathcal{T})=a\mathcal{T}^\frac{1}{2}+b,
\end{equation}
where $a$ and $b$ represent the integration constants.

We utilize Eqs.\eqref{39} and \eqref{42}, in conjunction with the
constraint on $f_{2}(\mathcal{T})$ as given in \eqref{47}, to
formulate a reconstruction framework employing the correspondence
approach. The differential equation for $f_{1}(\mathcal{Q})$ is
expressed as
\begin{equation}\label{48}
\frac{f_{1}(\mathcal{Q})}{2}-\mathcal{Q}
f_{1\mathcal{Q}}+a\mathcal{T}^\frac{1}{2}+\frac{1}{2}b=\alpha
H+\beta H^{2}.
\end{equation}
We use the power-law solution for the scale factor given as follows
\begin{equation}\label{49}
\mathrm{a}(t)=\mathrm{a}_{0}t^m,   \quad m > 0.
\end{equation}
In this context, $\mathrm{a}_{0}$ denotes the current value of the
scale factor. Employing this relation, the expressions for $H$, its
derivative, and the non-metricity scalar in terms of cosmic time $t$
are given as follows
\begin{equation*}
H=\frac{m}{t}, \quad\dot{H}=-\frac{m}{t^{2}}, \quad \mathcal{Q}=6
\frac{m^2}{t^2}.
\end{equation*}
Substituting \eqref{49} in \eqref{45}, it follows that
\begin{equation}\label{43-a}
\rho_{M}=d(\mathrm{a}_{0}t^m)^{-3},
\end{equation}
where $\rho_0=d$ for the sake of similicity. Using Eq.\eqref{49} in
\eqref{48}, we can find the function $f_{1}(\mathcal{Q})$ as
\begin{align}\nonumber
f_{1}(\mathcal{Q})&=\sqrt{\mathcal{Q}} \bigg[-\frac{a \sqrt{d}
2^{\frac{3 m}{8}+1} 3^{\frac{3 m}{8}}
\bigg(\frac{\mathcal{Q}}{m^2}\bigg)^{-\frac{3 m}{8}}}{\bigg(-\frac{3
m}{8}-\frac{1}{2}\bigg) \sqrt{\mathcal{Q}}}+\frac{2
b}{\sqrt{\mathcal{Q}}}+\frac{\beta m \mathcal{Q}}{\sqrt{6}}+\frac{2\
2^{3/4} \alpha \mathcal{Q}}{3 \sqrt[4]{3}}\bigg] \\\label{50} &+c_1
\sqrt{\mathcal{Q}},
\end{align}
where $c_{1}$ is the integration constant. Consequently, the
reconstructed $f\mathcal{(Q,T)}$ model is obtained by substituting
Eqs.\eqref{47} and \eqref{50} into \eqref{40} as follows
\begin{align}\nonumber
f(\mathcal{Q,T})&=\frac{1}{18} \bigg[-\frac{a \sqrt{d} 6^{\frac{3
m}{8}+2} \bigg(\frac{\mathcal{Q}}{m^2}\bigg)^{-\frac{3
m}{8}}}{-\frac{3 m}{8}-\frac{1}{2}}+18 \sqrt{\text{ad}}+54 b+18 c_1
\sqrt{\mathcal{Q}}
\\\label{51}
&+3 \sqrt{6} \beta  m \mathcal{Q}^{3/2}+4\ 6^{3/4} \alpha
\mathcal{Q}^{3/2}\bigg].
\end{align}

Now we write down this function in terms of redshift parameter $z$.
The expression for the deceleration parameter is given by
\begin{equation}\label{67}
q=-\frac{a\ddot{\mathrm{a}}}{\dot{\mathrm{a}}^{2}}=-1+\frac{1}{m}.
\end{equation}
In relation to the deceleration parameter, the evolution of the
cosmic scale factor can be characterized as
\begin{equation}\label{68}
\mathrm{a}(t)=t^{\frac{1}{(1+q)}},
\end{equation}
Here, we assume $\mathrm{a}_0$ to be equal to unity. It is worth
noting that the power-law model corresponds to an expanding universe
when $q > -1$. The description of both the expanding phase and the
present cosmic evolution is given by
\begin{equation}\label{69}
H=(1+q)^{-1}t^{-1}, \quad H_{0}=(1+q)^{-1}t^{-1}_{0}.
\end{equation}
In power-law cosmology, the evolution of the universe expansion is
determined by two fundamental parameters, the Hubble constant
($H_{0}$) and the deceleration parameter ($q$). By examining the
correlation between the scale factor $a$ and redshift $z$, we can
elucidate
\begin{equation}\label{70}
H=H_{0}\Gamma^{1+q},
\end{equation}
where $\Gamma=1+z$. Using Eq.\eqref{70} in \eqref{86}, $\mathcal{Q}$
appears as
\begin{align}\label{71}
\mathcal{Q}=6H_{0}^{2}\Gamma^{2+2q}.
\end{align}
The reconstructed model for the GGDE $f(\mathcal{Q,T})$ in terms of
the redshift parameter is derived by substituting this value in
Eq.\eqref{51}, resulting in
\begin{align}\nonumber
f\mathcal{(Q,T)}&=\frac{16 a \sqrt{d} \bigg(\frac{H_{0}^2 \Gamma^{2
q+2}}{m^2}\bigg)^{-\frac{3 m}{8}}}{3 m+4}+\sqrt{\text{ad}}+3
b+\sqrt{6} c_1 \sqrt{H_{0}^2 \Gamma^{2 q+2}}
\\\label{52}
&+\big(H_{0}^2 \Gamma^{2 q+2}\big)^{3/2}6 \beta  m +8 \sqrt[4]{6}
\alpha  \big(H_{0}^2 \Gamma^{2 q+2}\big)^{3/2}.
\end{align}
\begin{figure}\center
\epsfig{file=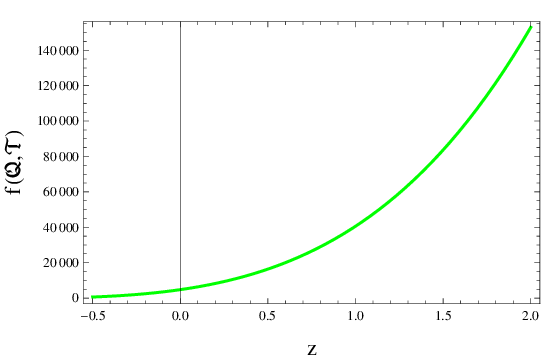,width=.7\linewidth} \caption{Plot of GGDE
$f\mathcal{(Q,T)}$ model against $z$.}
\end{figure}

For the graphical analysis, we set $a = 1$, $b = -4$, and $c_1 =
-15$. Figure \textbf{1} indicates that the reconstructed GGDE model
remains positive and increases with respect to $z$. The GGDE model
indicates a rapid expansion, according to this analysis. Figure
\textbf{2} illustrates the behavior of $\rho_{GGDE}$ and $P_{GGDE}$
with the redshift parameter. The energy density $\rho_{GGDE}$ is
positive and increasing, whereas $P_{GGDE}$ is negative and
consistent with the DE behavior. We then investigate the properties
of $\rho_{GGDE}$ and $P_{GGDE}$ in the context of the reconstructed
GGDE $f(\mathcal{Q,T})$ gravity model. Putting \eqref{52} into
\eqref{42} and \eqref{43}, we have
\begin{figure}\center
\epsfig{file=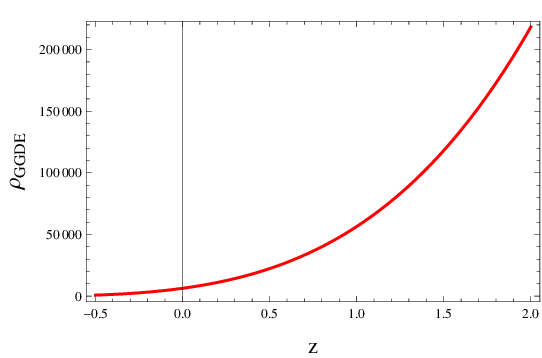,width=.5\linewidth}\epsfig{file=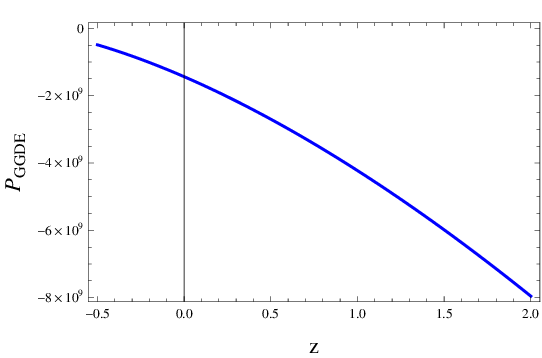,width=.5\linewidth}
\caption{Plots of $\rho_{GGDE}$ against $z$ (left) and $P_{GGDE}$
against $z$ (right).}
\end{figure}
\begin{align}\nonumber \rho_{GGDE}&=\frac{1}{18}
\bigg(\frac{H_{0}^2 \Gamma^{2 q+2}}{m^2}\bigg)^{-\frac{3 m}{8}}
\bigg[36 a \sqrt{d}-18 \bigg(\frac{H_{0}^2 \Gamma^{2
q+2}}{m^2}\bigg)^{\frac{3 m}{8}} \bigg\{\sqrt{\text{ad}} (d-1)
\\\nonumber
&+b (d-2)+2 \big(H_{0}^2 \Gamma^{2 q+2}\big)^{3/2} \big(4
\sqrt[4]{6} \alpha +3 \beta m\big)\bigg\}\bigg],
\\\nonumber
P_{GGDE}&=\frac{1}{36} \bigg[\frac{1}{H_{0}}\bigg\{(3 H_{0}-1)
\bigg\{6 \sqrt{6} \sqrt{H_{0}^2 \Gamma^{2 q+2}} \bigg\{c_1+3
\sqrt{6} \beta  H_{0}^2 m \Gamma^{2 q+2}
\\\nonumber
&+4\ 6^{3/4} \alpha H_{0}^2 \Gamma^{2 q+2}\bigg\}-\frac{72 a
\sqrt{d} m \bigg(\frac{H_{0}^2 \Gamma^{2 q+2}}{m^2}\bigg)^{-\frac{3
m}{8}}}{3 m+4}\bigg\}\bigg\}
\\\nonumber
&+\frac{1}{12 H_{0}^3}\bigg[\Gamma^{-3 (q+1)} \bigg\{\bigg\{18 a
\sqrt{d} m(3 m+8)\bigg(\frac{H_{0}^2 \Gamma^{2
q+2}}{m^2}\bigg)^{-\frac{3 m}{8}}\bigg\}(3 m+4)
\\\nonumber
&+\sqrt{6} \sqrt{H_{0}^2 \Gamma^{2 q+2}} \bigg\{-6 c_1+18 \sqrt{6}
\beta H_{0}^2 m \Gamma^{2 q+2}+24\ 6^{3/4} \alpha  H_{0}^2 \Gamma^{2
q+2}\bigg\}\bigg\}\bigg]\\\nonumber &-\frac{288 a \sqrt{d}
\bigg(\frac{H_{0}^2 \Gamma^{2 q+2}}{m^2}\bigg)^{-\frac{3 m}{8}}}{3
m+4} -18 \sqrt{\text{ad}}-54 b-18 \sqrt{6} c_1 \sqrt{H_{0}^2
\Gamma^{2 q+2}}
\\\nonumber
&-108 \beta  m \big(H_{0}^2 \Gamma^{2 q+2}\big)^{3/2}-144
\sqrt[4]{6} \alpha_{0}a \big(H_{0}^2 \Gamma^{2
q+2}\big)^{3/2}\bigg].
\end{align}

\section{Cosmological Analysis}

In this section, we explore the evolutionary stages of the universe
through different phases. To achieve this, we utilize the
reconstructed GGDE $f\mathcal{(Q,T)}$ model under non-interacting
conditions, as defined in Eq.\eqref{52}. Furthermore, we depict the
dynamics of various cosmological parameters, including the EoS
parameter, $\omega_{GGDE}-\omega'_{GGDE}$ and state finder planes.
The stability of this model is also examined.

\subsection{Equation of State Parameter}

The EoS parameter ($\omega=\frac{P}{\rho}$) of DE plays a crucial
role in describing both the cosmic inflationary phase and the
subsequent expansion of the universe. We analyze the criterion for
an accelerating universe, which occurs when the EoS parameter
$\omega<-\frac{1}{3}$. When $\omega=-1$, it corresponds to the
cosmological constant. However, when $\omega=\frac{1}{3}$ and
$\omega=0$, it signifies the radiation-dominated and
matter-dominated universe, respectively. Moreover, the phantom
scenario manifests when we assume $\omega<-1$. The expression for
the EoS parameter is given by
\begin{equation}\label{53}
\omega_{GGDE}=
\frac{P_{eff}}{\rho_{eff}}=\frac{P_{GGDE}}{\rho_{GGDE}+\rho_{M}}.
\end{equation}
Equations \eqref{42}, \eqref{43} and \eqref{43-a} are employed in
the aforementioned expression to compute the respective parameter as
\begin{align}\nonumber
\omega_{GGDE}&=-\bigg[f^3 \Gamma^{-2 q-2} \bigg\{a \sqrt{d}
2^{\frac{3 m}{8}+1} 3^{\frac{3 m}{8}+2} \big(-432 h^4 m \Gamma^{4
q+4} -48 h^2 \Gamma^{2 q+2}\big(12 h^2
\\\nonumber
&\times\Gamma^{2 q+2}-3 h m \Gamma^{2 q+2}\big) +3 h m (3 m+8)
\Gamma^{q+1}\big) -6^{\frac{3 mx}{8} +\frac{1}{2}} (3 m+4) \sqrt{h^2
\Gamma^{2 q+2}}
\\\nonumber
&\times \bigg(\frac{h^2 \Gamma^{2 q+2}}{m^2}\bigg)^{\frac{3 m}{8}}
\bigg\{108 \sqrt{6} \sqrt{\text{ad}} \big(H_{0}^2 \Gamma^{2
q+2}\big)^{3/2}+324 \sqrt{6} b \big(H_{0}^2 \Gamma^{2
q+2}\big)^{3/2}+216
\\\nonumber
&\times c_1 H_{0}^3 \Gamma^{4 q+4}+18 c_1 H_{0} \Gamma^{q+1}-1296
\sqrt{6} \beta  H_{0}^6 m \Gamma^{6 q+6} -1728\ 6^{3/4} \alpha
H_{0}^6 \Gamma^{6 q+6}
\\\nonumber
&+648 \sqrt{6} \beta H_{0}^5 m \Gamma^{6 q+6} +864\ 6^{3/4} \alpha
H_{0}^5 \Gamma^{6 q+6}-54 \sqrt{6} \beta H_{0}^3 m \Gamma^{3
q+3}-72\ 6^{3/4}
\\\nonumber
&\times \alpha  H_{0}^3 \Gamma^{3 q+3}\bigg\}\bigg\}\bigg] \bigg[18
H_{0}^2 (3 m+4) \bigg\{a \sqrt{d} f^3 2^{\frac{3 m}{8}} 3^{\frac{3
m}{8}+2} m +H_{0}^2 6^{\frac{3 m}{8}+1}\Gamma^{2 q+2}
\\\nonumber
&\times \bigg(\frac{H_{0}^2 \Gamma^{2 q+2}}{m^2}\bigg)^{\frac{3
m}{8}} \big(6 f^3 \sqrt{H_{0}^2 \Gamma^{2 q+2}} \big(4 \sqrt[4]{6}
\alpha +3 \beta m\big)-12 d\big)\bigg\}\bigg]^{-1}.
\end{align}
Figure \textbf{3} illustrates the behavior of EoS parameter against
$z$ from which one can find that phantom epoch for current as well
as late time cosmic evolution.
\begin{figure}\center
\epsfig{file=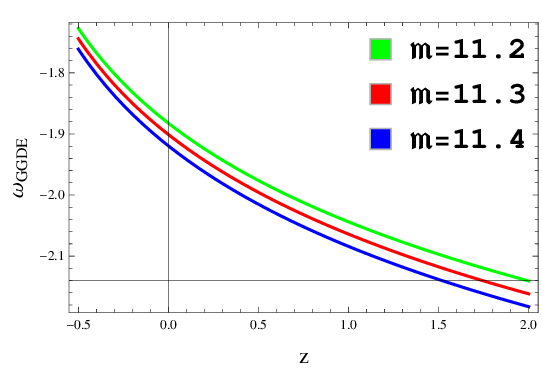,width=.6\linewidth}\caption{Plot of EoS
parameter against $z$.}
\end{figure}

\subsection{$\omega_{GGDE}-\omega'_{GGDE}$ Plane}

Here, we utilize the $\omega_{GGDE}-\omega'_{GGDE}$ phase plane,
where $\omega'_{GGDE}$ indicates the evolutionary mode of
$\omega_{GGDE}$ and prime denotes the derivative with respect to
$\mathcal{Q}$. This cosmic plane was established by Caldwell and
Linder \cite{32} to investigate the quintessence DE paradigm which
can be split into freezing $(\omega_{GGDE} <0,~\omega'_{GGDE}<0 )$
and thawing $(\omega_{GGDE}<0,~\omega'_{GGDE}>0 )$ regions. To
depict the prevailing cosmic expansion paradigm, the freezing region
signifies a more accelerated phase compared to thawing. The cosmic
trajectories of $\omega_{GGDE}-\omega'_{GGDE}$ plane for specific
choices of $m$ are shown in Figure \textbf{4} which provides the
freezing region of the cosmos. The expression of $\omega'_{GGDE}$ is
given in Appendix \textbf{A}.
\begin{figure}\center
\epsfig{file=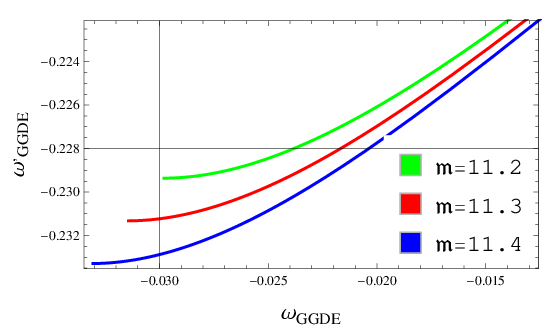,width=.6\linewidth}\caption{Plot of
$\omega_{GGDE}-\omega'_{GGDE}$ versus $z$.}
\end{figure}

\subsection{State Finder Analysis}

One of the techniques to examine the dynamics of the cosmos using a
cosmological perspective is state finder analysis. It is an
essential approach for understanding numerous DE models. As a
combination of the Hubble parameter and its temporal derivatives,
Sahni et al. \cite{33-a} established two dimensionless parameters
$(r,s)$ given as
\begin{equation}\label{64}
r=\frac{\dddot{a}}{a H^{3}}, \quad s=\frac{r-1}{3(q-\frac{1}{2})}.
\end{equation}
The acceleration of cosmic expansion is determined by the parameter
$s$, while the deviation from pure power-law behavior is precisely
described by the parameter $r$. This is a geometric diagnostic that
does not favor any particular cosmological paradigm. It is such an
approach that does not depend upon any specific model to distinguish
between numerous DE scenarios, i.e., CG (Chaplygin gas), HDE
(Holographic DE), SCDM (standard CDM) and quintessence.

Several DE scenarios for the appropriate choices of $r$ and $s$
parametric values are given below.
\begin{itemize}
\item When $r=1$, $s=0$, it indicates the CDM model.
\item If $r=1$, $s=1$, then it denotes SCDM epoch.
\item When $r=1$, $s=\frac{2}{3}$, this epoch demonstrates the HDE
model.
\item When we have $r>1$, $s<0$, we get CG scenario.
\item Lastly, $r<1$, $s>0$ corresponds to quintessence paradigm.
\end{itemize}
For our considered setup, the parameters $r$ and $s$ in terms of
required factors are given in Appendix \textbf{B}. The graphical
analysis of $r-s$ phase plane in Figure \textbf{5} gives $r>1$ and
$s<0$, indicating the CG model.
\begin{figure}\center
\epsfig{file=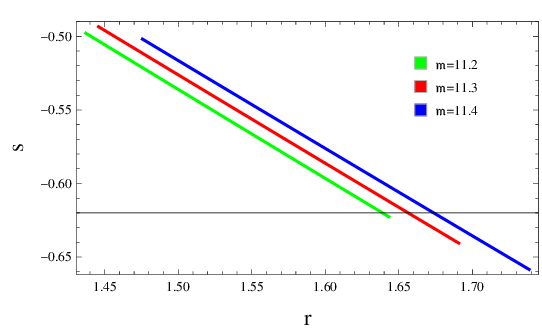,width=.7\linewidth}\caption{Plot of $r-s$ plane
against $z$.}
\end{figure}

\subsection{Squared Speed of Sound Parameter}

Perturbation theory provides a direct analysis for evaluating the
stability of the DE model through examination of the sign of
$(\nu_{s}^{2})$. In this context, we investigate the squared speed
of sound parameter to analyze the stability of the GGDE
$f(\mathcal{Q,T})$ model, represented by
\begin{equation}\label{55}
\nu_{s}^{2}=\frac{P_{GGDE}}{\rho'_{GGDE}}\omega'_{GGDE}
+\omega_{GGDE}.
\end{equation}
A positive value signifies a stable configuration, whereas a
negative value indicates unstable behavior for the associated model.
Substituting the necessary expressions on the right-hand side of the
equation above for the reconstructed model, we derive the squared
speed of sound parameter as provided in Appendix \textbf{C}. Figure
\textbf{6} illustrates that the speed component remains positive for
all assumed values of $m$, indicating the stability of the
reconstructed GGDE $f(\mathcal{Q,T})$ model throughout the cosmic
evolution.
\begin{figure}\center
\epsfig{file=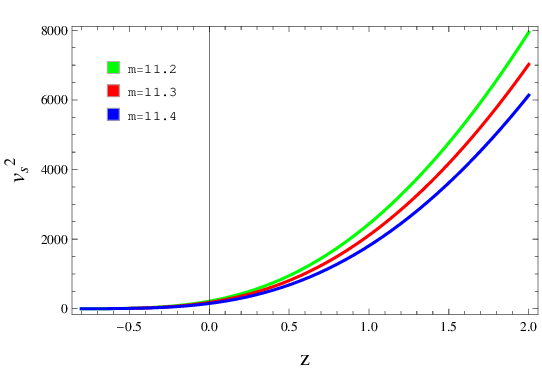,width=.7\linewidth}\caption{Plot of
$\nu_{s}^{2}$ against $z$.}
\end{figure}

\section{Final Remarks}

In this study, we have explored various features of the GGDE model
within the framework of the recently developed $f\mathcal{(Q,T)}$
theory of gravity. To comprehend the influence of $\mathcal{Q}$ and
$\mathcal{T}$ in the GGDE ansatz, we have adopted a specific
$f(\mathcal{Q,T})=f_{1}(\mathcal{Q})+f_{2}(\mathcal{T})$ model. We
have formulated a reconstruction framework involving a power-law
scale factor and a corresponding scenario for the flat FRW universe.
Subsequently, we have analyzed the EoS parameter,
$\omega_{GGDE}-\omega'_{GGDE}$ and state finder planes, and
conducted a squared sound speed analysis for the derived model. We
summarize the main results as follows.
\begin{itemize}
\item In the non-interacting case, the reconstructed GGDE $f(\mathcal{Q,T})$
model shows an increasing trend for $z$, indicating the realistic
nature of the reconstructed model (Figure \textbf{1}).
\item The energy density demonstrates positive behavior, showing an
increase, while the pressure exhibits negative behavior. These
patterns align with the characteristics of DE (Figure \textbf{2}).
\item
The early and late time universe are characterized by the EoS
parameter, which involves phantom field and DE. Additionally, we
have noted a trend where this parameter assumes increasingly
negative values below -1 (Figure \textbf{3}). These observations
align with the current understanding of accelerated cosmic behavior.
\item
The evolutionary pattern in the $\omega_{GGDE}-\omega'_{GGDE}$ plane
indicates a freezing region for all values of $m$ (Figure
\textbf{4}). This affirms that the non-interacting GGDE
$f(\mathcal{Q,T})$ gravity model implies a more accelerated
expanding universe.
\item The $r-s$ plane depicts the CG model as both $r$
and $s$ satisfy the respective model criteria (Figure \textbf{5}).
\item We have found that the squared speed of sound parameter is
positive and hence GGDE $f(\mathcal{Q,T})$ gravity model is stable
for all values of $m$ (Figure \textbf{6}).
\end{itemize}

Finally, we conclude that our results align with the current
observational data \cite{34-a}, as provided below
\begin{eqnarray}\nonumber
\omega_{GGDE}&=&-1.023^{+0.091}_{-0.096}\quad(\text{Planck
TT+LowP+ext}),\\\nonumber
\omega_{GGDE}&=&-1.006^{+0.085}_{-0.091}\quad(\text{Planck
TT+LowP+lensing+ext}),\\\nonumber
\omega_{GGDE}&=&-1.019^{+0.075}_{-0.080}\quad (\text{Planck TT, TE,
EE+LowP+ext}).
\end{eqnarray}
These values have been established using diverse observational
methodologies, ensuring a confidence level of 95$\%$. Moreover, we
have verified that the cosmic diagnostic state finder parameters for
our derived model align with the constraints and limitations on the
kinematics of the universe \cite{35}. Sharif and Saba \cite {62}
established the correspondence of modified Gauss-Bonnet theory with
GGDE paradigm and found phantom phase as well as the stable
reconstructed model in non-interacting case. Our findings align with
these results. Our results are also consistent with the recent work
in $f\mathcal{(Q)}$ theory \cite{61}.

\section*{Appendix A: Calculation of $\omega'_{GGDE}$}
\renewcommand{\theequation}{A\arabic{equation}}
\setcounter{equation}{0}

\begin{align}\nonumber
\omega'_{GGDE}&=\bigg[f^3 \Gamma^{-4 q-4} \bigg\{2^{\frac{3 m}{8}+1}
3^{\frac{3 m}{8}+2} a \sqrt{d} \big(3 H_{0} m (3 m+8)
\Gamma^{q+1}-48 H_{0}^2 \big(12 H_{0}^2 \Gamma^{2 q+2}
\\\nonumber&
-3 H_{0} m \Gamma^{2 q+2}\big) \Gamma^{2 q+2}-432 H_{0}^4 m
\Gamma^{4 q+4}\big) -6^{\frac{3 m}{8}+\frac{1}{2}} (3 m+4)
\sqrt{H_{0}^2 \Gamma^{2 q+2}}
\\\nonumber&
\times \bigg(\frac{H_{0}^2 \Gamma^{2 q+2}}{m^2}\bigg)^{\frac{3
m}{8}}\bigg\{324 \sqrt{6} b \big(H_{0}^2 \Gamma^{2
q+2}\big)^{3/2}+108 \sqrt{6} \sqrt{\text{ad}} \big(H_{0}^2 \Gamma^{2
q+2}\big)^{3/2}
\\\nonumber
&-726^{3/4} H_{0}^3 \Gamma^{3 q+3} \alpha -1728\ 6^{3/4} H_{0}^6
\Gamma^{6 q+6} \alpha+864\ 6^{3/4} H_{0}^5 \Gamma^{6 q+6} \alpha
\\\nonumber
&-54 \sqrt{6} H_{0}^3m \Gamma^{3 q+3} \beta -1296 \sqrt{6} H_{0}^6m
\Gamma^{6 q+6} \beta+648 \sqrt{6} H_{0}^5 m \Gamma^{6 q+6} \beta
\\\nonumber
&+18 H_{0} \Gamma^{q+1} c_1+216 H_{0}^3 \Gamma^{4 q+4}
c_1\bigg\}\bigg\}\bigg]\bigg[108 H_{0}^4 (3 m+4) \bigg\{6^{\frac{3
m}{8}+1} H_{0}^2 \Gamma^{2 q+2}
\\\nonumber
&\times\big(6 f^3 \sqrt{H_{0}^2 \Gamma^{2 q+2}}\big(4 \sqrt[4]{6}
\alpha +3 m \beta \big) -12 d\big) \bigg(\frac{H_{0}^2 \Gamma^{2
q+2}}{m^2}\bigg)^{\frac{3 m}{8}} +2^{\frac{3 m}{8}} 3^{\frac{3
m}{8}+2}
\\\nonumber
&\times a \sqrt{d} f^3 m\bigg\}\bigg]^{-1}+\bigg[f^3 \Gamma^{-2 q-2}
\bigg\{2^{\frac{3 m}{8}} 3^{\frac{3 m}{8}+1} f^3 \sqrt{H_{0}^2
\Gamma^{2 q+2}} \big(4 \sqrt[4]{6} \alpha+3 m \beta \big)
\\\nonumber
&\times\bigg(\frac{H_{0}^2 \Gamma^{2 q+2}}{m^2}\bigg)^{\frac{3
m}{8}} +6^{\frac{3 m}{8}}\big(6 f^3 \sqrt{H_{0}^2 \Gamma^{2 q+2}}
\big(4 \sqrt[4]{6} \alpha +3 m \beta \big) \bigg(\frac{H_{0}^2
\Gamma^{2 q+2}}{m^2}\bigg)^{\frac{3 m}{8}}
\\\nonumber
&+\bigg[2^{\frac{3 m}{8}-3} 3^{\frac{3 m}{8}+1}H_{0}^2 \Gamma^{2
q+2} \big(6 f^3 \sqrt{H_{0}^2 \Gamma^{2 q+2}} \big(4 \sqrt[4]{6}
\alpha+3 m \beta \big)\bigg(\frac{H_{0}^2 \Gamma^{2 q+2}}{m^2}
\bigg)^{\frac{3 m}{8}-1}\bigg]
\\\nonumber
&\times\frac{1}{m}\bigg\}\bigg(2^{\frac{3 m}{8}+1} 3^{\frac{3
m}{8}+2}a \sqrt{d} \big\{3 H_{0} m (3 m+8) \Gamma^{q+1}-48 H_{0}^2
\big(12 H_{0}^2 \Gamma^{2 q+2}
\\\nonumber
&-3 H_{0} m \Gamma^{2 q+2}\big) \Gamma^{2 q+2}-432 H_{0}^4 m
\Gamma^{4 q+4}\big\} -6^{\frac{3 m}{8}+\frac{1}{2}} (3
m+4)\bigg(\frac{H_{0}^2 \Gamma^{2 q+2}}{m^2}\bigg)^{\frac{3 m}{8}}
\\\nonumber
&\times\sqrt{H_{0}^2 \Gamma^{2 q+2}} \big(324 \sqrt{6} b
\times\big(H_{0}^2 \Gamma^{2 q+2}\big)^{3/2}+108 \sqrt{6}
\sqrt{\text{ad}} \big(H_{0}^2 \Gamma^{2 q+2}\big)^{3/2}
\\\nonumber
& -726^{3/4} H_{0}^3 \Gamma^{3 q+3} \alpha -1728 \ 6^{3/4} H_{0}^6
\Gamma^{6 q+6} \alpha +864\ 6^{3/4} H_{0}^5 \Gamma^{6 q+6} \alpha
\\\nonumber
&-54 \sqrt{6} H_{0}^3 m \Gamma^{3 q+3} \beta -1296 \sqrt{6} H_{0}^6
m \Gamma^{6 q+6} \beta +648 \sqrt{6} H_{0}^5m\Gamma^{6 q+6} \beta
\\\nonumber
&+18 H_{0} \Gamma^{q+1} c_1+216 H_{0}^3 \Gamma^{4 q+4}
c_1\bigg)\bigg)\bigg] \bigg[18 H_{0}^2 (3 m+4) \bigg(6^{\frac{3
m}{8} +1} H_{0}^2 \Gamma^{2 q+2}
\\\nonumber
&\times \big(6 f^3 \sqrt{H_{0}^2 \Gamma^{2 q+2}} \big(4\sqrt[4]{6}
\alpha +3 m \beta \big)-12 d\big) \bigg(\frac{H_{0}^2 \Gamma^{2
q+2}}{m^2}\bigg)^{\frac{3 m}{8}} +2^{\frac{3 m}{8}} 3^{\frac{3
m}{8}+2}
\\\nonumber
&\times a \sqrt{d} f^3 m\bigg)^2\bigg]^{-1}-\bigg[f^3\Gamma^{-2
q-2}\big(-6^{\frac{3 m}{8}+\frac{1}{2}} (3 m+4)\sqrt{H_{0}^2
\Gamma^{2 q+2}} \bigg\{-12 6^{3/4}
\\\nonumber
&\times H_{0} \alpha \Gamma^{q+1}-9 \sqrt{6} H_{0} m \beta
\Gamma^{q+1} +108 H_{0}^2 c_1 \Gamma^{2 q+2}+36 H_{0} c_1 \Gamma^{2
q+2}
\\\nonumber
&-432\ 6^{3/4} H_{0}^4 \alpha \Gamma^{4 q+4}+2886^{3/4} H_{0}^3
\alpha \Gamma^{4 q+4} -324 \times \sqrt{6} H_{0}^4 m \beta \Gamma^{4
q+4}
\\\nonumber
&+216 \sqrt{6} H_{0}^3 m \beta \Gamma^{4 q+4}+81 \sqrt{6} b
\sqrt{H_{0}^2 \Gamma^{2 q+2}} +27 \sqrt{6}
\sqrt{\text{ad}}\sqrt{H_{0}^2 \Gamma^{2 q+2}}\big\}
\\\nonumber
&\times \bigg(\frac{H_{0}^2 \Gamma^{2 q+2}}{m^2}\bigg)^{\frac{3
m}{8}} +2^{\frac{3 m}{8}+1} 3^{\frac{3 m}{8}+2} a \sqrt{d}
\bigg\{-96 H_{0}^2 \Gamma^{2 q+2} -72 H_{0}^2 m \Gamma^{2 q+2}
\\\nonumber
&-8 \big(12 H_{0}^2 \Gamma^{2 q+2}-3 H_{0} m \Gamma^{2
q+2}\big)\bigg\}-\bigg[2^{\frac{3 m}{8}-\frac{3}{2}} 3^{\frac{3
m}{8}-\frac{1}{2}} (3 m+4)\bigg(\frac{H_{0}^2 \Gamma^{2
q+2}}{m^2}\bigg)^{\frac{3 m}{8}}
\\\nonumber
&\times\{324 \sqrt{6} b \big(H_{0}^2 \Gamma^{2 q+2}\big)^{3/2}
+108\sqrt{6} \sqrt{\text{ad}} \big(H_{0}^2 \Gamma^{2 q+2}\big)^{3/2}
-6^{3/4} H_{0}^3 \Gamma^{3 q+3} \alpha
\\\nonumber
&-1728\ 6^{3/4} H_{0}^6 \Gamma^{6 q+6} \alpha+864\ 6^{3/4}
H_{0}^5\Gamma^{6 q+6} \alpha -54 \sqrt{6} H_{0}^3 m \Gamma^{3 q+3}
\beta
\\\nonumber
&-1296 \sqrt{6} H_{0}^6 m \Gamma^{6 q+6} \beta +\sqrt{6} H_{0}^5
m\Gamma^{6 q+6} \beta +18 H_{0} \Gamma^{q+1} c_1 +H_{0}^3 \Gamma^{4
q+4} c_1\bigg\}\bigg]
\\\nonumber
&\times\bigg[\sqrt{H_{0}^2 \Gamma^{2 q+2}}\bigg]^{-1}
-\bigg[2^{\frac{3 m}{8}-\frac{7}{2}} 3^{\frac{3 m}{8}+\frac{1}{2}}
(3 m+4) \sqrt{H_{0}^2 \Gamma^{2 q+2}} \bigg(\frac{H_{0}^2 \Gamma^{2
q+2}}{m^2}\bigg)^{\frac{3 m}{8}-1}
\\\nonumber
&\times\big(324 \sqrt{6} b \big(H_{0}^2 \Gamma^{2
q+2}\big)^{3/2}+108 \sqrt{6}\big(H_{0}^2 \Gamma^{2 q+2}\big)^{3/2}
-72\ 6^{3/4} H_{0}^3 \Gamma^{3 q+3} \alpha
\\\nonumber
&-1728\ 6^{3/4} H_{0}^6 \Gamma^{6 q+6} \alpha +864\ 6^{3/4}
H_{0}^5\Gamma^{6 q+6} \alpha -54 \sqrt{6} H_{0}^3 m \Gamma^{3 q+3}
\beta
\\\nonumber
&-1296 \sqrt{6} H_{0}^6 m \Gamma^{6 q+6} \beta +648 \sqrt{6} H_{0}^5
m \Gamma^{6 q+6} \beta+18 H_{0} \Gamma^{q+1} c_1
\\\nonumber
&+216 H_{0}^3 \Gamma^{4 q+4} c_1\big)]\bigg[18 H_{0}^2 (3 m+4)
\big(6^{\frac{3 m}{8}+1} H_{0}^2 \Gamma^{2 q+2}\big(6 f^3
\sqrt{H_{0}^2 \Gamma^{2 q+2}}
\\\label{61}
&\times\big(4 \sqrt[4]{6} \alpha +3 m \beta \big)-12 d\big)
\bigg(\frac{H_{0}^2 \Gamma^{2 q+2}}{m^2}\bigg)^{\frac{3
m}{8}}+2^{\frac{3 m}{8}} 3^{\frac{3 m}{8}+2} a \sqrt{d} f^3
m\big)\bigg].
\end{align}

\section*{Appendix B: Evaluation of $r$ and $s$ Parameters}
\renewcommand{\theequation}{B\arabic{equation}}
\setcounter{equation}{0}

\begin{align}\nonumber
r&=2 \bigg(\frac{3}{2}\bigg[f^3 \Gamma^{-4 q-4} \big(2^{\frac{3
m}{8}+1} 3^{\frac{3 m}{8}+2} a \sqrt{d} \big\{3 H_{0} m (3 m+8)
\Gamma^{q+1}-48 H_{0}^2 \big(12 H_{0}^2 \Gamma^{2 q+2}
\\\nonumber
& - \Gamma^{2 q+2}\big) \Gamma^{2 q+2} -432 H_{0}^4 m \Gamma^{4
q+4}\big\}-6^{\frac{3 m}{8} +\frac{1}{2}} (3 m+4) \sqrt{^H_{0}2
\Gamma^{2 q+2}} \bigg(\frac{h^2 \Gamma^{2 q+2}}{m^2}\bigg)^{\frac{3
m}{8}}
\\\nonumber
&\times \bigg(324 \sqrt{6} b \big(H_{0}^2 \Gamma ^{2 q+2}\big)^{3/2}
+108 \sqrt{6} \sqrt{\text{ad}} \big(H_{0}^2 \Gamma^{2
q+2}\big)^{3/2}-72\ 6^{3/4} h^3 \Gamma^{3 q+3} \alpha
\\\nonumber
&\times\ 6^{3/4} H_{0}^6 \Gamma^{6 q+6} \alpha +864\ 6^{3/4} h^5
\Gamma^{6 q+6} \times\alpha -54 \sqrt{6} H_{0}^3 m \Gamma^{3 q+3}
\beta -1296 \sqrt{6} H_{0}^6m
\\\nonumber
&\times\Gamma^{6 q+6}  \beta +648 \sqrt{6} H_{0}^5 m \Gamma^{6 q+6}
\beta +18 H_{0} \Gamma^{q+1} c_1+216 h^3 \Gamma^{4 q+4}
c_1\bigg)\bigg)\bigg] \bigg[108 h^4
\\\nonumber
&\times \big(6^{\frac{3 m}{8}+1}H_{0}^2 \Gamma^{2 q+2} \big(6 f^3
\sqrt{H_{0}^2 \Gamma^{2 q+2}} \bigg\{4 \sqrt[4]{6} \alpha +3 m \beta
\bigg\}\bigg(\frac{H_{0}^2 \Gamma^{2 q+2}}{m^2}\bigg)^{\frac{3
m}{8}}
\\\nonumber
&\times 3^{\frac{3 m}{8}+2} a \sqrt{d} f^3 m\bigg)\bigg]^{-1}
+\bigg[f^3 \Gamma^{-2 q-2} \bigg(2^{\frac{3 m}{8}} 3^{\frac{3
m}{8}+1} f^3 \sqrt{H_{0}^2 \Gamma^{2 q+2}} \big(4 \sqrt[4]{6} \alpha
r
\\\nonumber
&+3 m \beta \big)\bigg(\frac{H_{0}^2 \Gamma^{2
q+2}}{m^2}\bigg)^{\frac{3 m}{8}} +6^{\frac{3 m}{8}} \bigg(6 f^3
\sqrt{H_{0}^2 \Gamma^{2 q+2}} \bigg(4\sqrt[4]{6} \alpha +3 m \beta
\bigg)
\\\nonumber
&-12 d\bigg) \bigg(\frac{H_{0}^2 \Gamma^{2 q+2}}{m^2}\bigg)^{\frac{3
m}{8}}+\frac{1}{m}\bigg[2^{\frac{3 m}{8}-3} 3^{\frac{3 m}{8}+1} h^2
\Gamma^{2 q+2} \big(6 f^3 \sqrt{h^2 \Gamma^{2 q+2}} \bigg(4
\sqrt[4]{6} \alpha
\\\nonumber
&+3 m \beta \bigg)- \bigg(\frac{h^2 \Gamma^{2
q+2}}{m^2}\bigg)^{\frac{3 m}{8}-1}\bigg]\bigg(2^{\frac{3 m}{8}+1}
3^{\frac{3 m}{8}+2} a \sqrt{d} \bigg(3 h m (3 m+8) \Gamma^{q+1}
\\\nonumber
& -48 h^2 \bigg(12 H_{0}^2 \Gamma^{2 q+2} -3 H_{0} m \Gamma^{2
q+2}\bigg)-432 h^4 m \Gamma^{4 q+4}\bigg) - \bigg(\frac{h^2
\Gamma^{2 q+2}}{m^2}\bigg)^{\frac{3 m}{8}}
\\\nonumber&
\times\sqrt{h^2 \Gamma^{2 q+2}} \big(324 \sqrt{6} b \big(h^2
\Gamma^{2 q+2}\bigg)^{3/2} 108 \sqrt{6} \sqrt{\text{ad}} + \bigg(h^2
\Gamma^{2 q+2}\bigg)^{3/2}-72\ 6^{3/4}
\\\nonumber
&\times h^3 \Gamma^{3 q+3} \alpha -1728\ 6^{3/4} h^6 \Gamma^{6 q+6}
\alpha +864\ 6^{3/4} h^5 \Gamma^{6 q+6} \alpha-54 \sqrt{6} h^3 m
\Gamma^{3 q+3} \beta
\\\nonumber
&-1296 \sqrt{6} H_{0}^6 m \Gamma^{6 q+6} \beta+216 H_{0}^3 \Gamma^{4
q+4} c_1\bigg]\bigg[18 h^2 (3 m+4) \bigg(6^{\frac{3 m}{8}+1} H_{0}^2
\Gamma^{2 q+2}
\\\nonumber
&\times 6 f^3 \sqrt{H_{0}^2 \Gamma^{2 q+2}} \big(4 \sqrt[4]{6}
\alpha 3 m \beta \big)-12 d\big) \bigg(\frac{H_{0}^2 \Gamma^{2
q+2}}{m^2}\bigg)^{\frac{3 m}{8}}+2^{\frac{3 m}{8}} 3^{\frac{3
m}{8}+2}f^3 m\bigg)^2\bigg]^{-1}
\\\nonumber
&-\bigg[f^3 \Gamma^{-2 q-2} \bigg(-6^{\frac{3 m}{8}+\frac{1}{2}}(3
m+4) \sqrt{H_{0}^2 \Gamma^{2 q+2}} \bigg(-12 6^{3/4} H_{0} \alpha
\Gamma^{q+1}
\\\nonumber
&+36 H_{0} c_1 \Gamma^{2 q+2}-432\ 6^{3/4} h^4 \alpha \Gamma^{4 q+4}
+288\ 6^{3/4} h^3 \alpha \Gamma^{4 q+4} -324 \sqrt{6} H_{0}^4 m
\\\nonumber
&\times\beta \Gamma^{4 q+4}+216 \sqrt{6} H_{0}^3 m \beta \Gamma^{4
q+4} + \sqrt{H_{0}^2 \Gamma^{2 q+2}} +27 \sqrt{6} \sqrt{\text{ad}}
\sqrt{H_{0}^2 \Gamma^{2 q+2}}\bigg)
\\\nonumber
&\times\bigg(\frac{H_{0}^2 \Gamma^{2 q+2}}{m^2}\bigg)^{\frac{3
m}{8}}+2^{\frac{3 m}{8}+1} 3^{\frac{3 m}{8}+2} a \sqrt{d} \bigg\{-
H_{0}^2 \Gamma^{2 q+2}- H_{0}^2 m \Gamma^{2 q+2}
\\\nonumber
&-8 \big(12 h^2 \Gamma^{2 q+2}-3 H_{0} m \Gamma^{2
q+2}\big)\bigg\}-\bigg[2^{\frac{3 m}{8}-\frac{3}{2}} 3^{\frac{3
m}{8}-\frac{1}{2}} (3 m+4) \bigg(\frac{H_{0}^2 \Gamma^{2
q+2}}{m^2}\bigg)^{\frac{3 m}{8}}
\\\nonumber
&\times\bigg(324 \sqrt{6} b \bigg(H_{0}^2 \Gamma^{2 q+2}\bigg)^{3/2}
+108 \sqrt{6}\sqrt{\text{ad}}\bigg(H_{0}^2 \Gamma^{2
q+2}\bigg)^{3/2} -72\ 6^{3/4}
\\\nonumber
&\times H_{0}^3 \Gamma^{3 q+3} \alpha -1728\ 6^{3/4} H_{0}^6
\Gamma^{6 q+6} \alpha +864\ 6^{3/4} H_{0}^5 \Gamma^{6 q+6} \alpha
-54 \sqrt{6}
\\\nonumber
&\times H_{0}^3 m \Gamma^{3 q+3} \beta - \sqrt{6} h^6 m \Gamma^{6
q+6} \beta + \sqrt{6} h^5 m \Gamma^{6 q+6} \beta +18 H_{0}
\Gamma^{q+1} c_1
\\\nonumber
&+216 h^3 \Gamma^{4 q+4} c_1\bigg)\bigg] \times\bigg[\sqrt{H_{0}^2
\Gamma^{2 q+2}}\bigg]^{-1}-\frac{1}{m}\bigg[2^{\frac{3 m}{8}
-\frac{7}{2}} 3^{\frac{3 m}{8}+\frac{1}{2}} \bigg(\frac{H_{0}^2
\Gamma^{2 q+2}}{m^2}\bigg)^{\frac{3 m}{8}-1}
\\\nonumber
&\times\sqrt{h^2 \Gamma^{2 q+2}} \bigg(324 \sqrt{6} b \bigg(H_{0}^2
\Gamma^{2 q+2}\bigg)^{3/2} +\sqrt{6} \sqrt{\text{ad}} \bigg(h^2
\Gamma^{2 q+2}\bigg)^{3/2}
\\\nonumber
&-72\ 6^{3/4} H_{0}^3 \Gamma^{3 q+3} \alpha -1728\ 6^{3/4} H_{0}^6
\Gamma^{6 q+6} \alpha +\ 6^{3/4} H_{0}^5 \Gamma^{6 q+6} \alpha -54
\sqrt{6} h^3
\\\nonumber
&\times m \Gamma^{3 q+3} \beta -1296 \sqrt{6} H_{0}^6 m \Gamma^{6
q+6} \beta +648 \sqrt{6} h^5 m \Gamma^{6 q+6} \beta +18 h
\Gamma^{q+1} c_1
\\\nonumber
&+216 h^3 \Gamma^{4 q+4} c_1\bigg)\bigg]\bigg]\bigg[18 h^2 (3 m+4)
\big(6^{\frac{3 m}{8}+1} h^2 \Gamma^{2 q+2} \big(6 f^3 \sqrt{h^2
\Gamma^{2 q+2}} \big(4 \sqrt[4]{6} \alpha
\\\nonumber
&+3 m \beta \big)-\big) \bigg(\frac{h^2 \Gamma^{2
q+2}}{m^2}\bigg)^{\frac{3 m}{8}}+2^{\frac{3 m}{8}} 3^{\frac{3
m}{8}+2} a \sqrt{d} f^3 m\bigg)\bigg]^{-1}\bigg)
+\frac{1}{2}\bigg)+\frac{3}{2}\bigg[f^3 \Gamma^{-4 q-4}
\\\nonumber
&\times\bigg(2^{\frac{3 m}{8}+1} 3^{\frac{3 m}{8}+2} a
\sqrt{d}\bigg(3 H_{0} m (3 m+8) \Gamma^{q+1} -48 H_{0}^2 \bigg(12
H_{0}^2 \Gamma^{2 q+2} -3 H_{0} m
\\\nonumber
&\times\Gamma^{2 q+2}\bigg) \Gamma^{2 q+2} -H_{0}^4 m \Gamma^{4
q+4}\bigg)-6^{\frac{3 m}{8} +\frac{1}{2}} (3 m+4) \sqrt{H_{0}^2
\Gamma^{2 q+2}} \bigg(\frac{H_{0}^2 \Gamma^{2
q+2}}{m^2}\bigg)^{\frac{3 m}{8}}
\\\nonumber
&\times\bigg(324 \sqrt{6} b \bigg(H_{0}^2 \Gamma^{2 q+2}\bigg)^{3/2}
+108 \sqrt{6}\sqrt{\text{ad}}  \bigg(H_{0}^2 \Gamma^{2
q+2}\bigg)^{3/2} -72\ 6^{3/4} H_{0}^3 \Gamma^{3 q+3} \alpha
\\\nonumber
& -1728\ 6^{3/4} H_{0}^6 \Gamma^{6 q+6} \alpha +864\ 6^{3/4} H_{0}^5
\Gamma^{6 q+6} \alpha -54 \sqrt{6} H_{0}^3 m \Gamma^{3 q+3} \beta
-1296 \sqrt{6} H_{0}^6 m
\\\nonumber
&\times \Gamma^{6 q+6} \beta +648 \sqrt{6} H_{0}^5 m \Gamma^{6 q+6}
\beta \Gamma^{q+1} c_1 +216 H_{0}^3 \Gamma^{4 q+4}
c_1\bigg)\bigg)\bigg] \bigg[108 H_{0}^4 (3 m+4)
\\\nonumber
&\times \big(6^{\frac{3 m}{8}+1} H_{0}^2 \Gamma^{2 q+2}
\big(\sqrt{H_{0}^2 \Gamma^{2 q+2}}\big(4 \sqrt[4]{6} \alpha +3 m
\beta \big) -12 d\big) \bigg(\frac{H_{0}^2 \Gamma^{2
q+2}}{m^2}\bigg)^{\frac{3 m}{8}}
\\\nonumber
&\times \bigg\{2^{\frac{3 m}{8}} 3^{\frac{3 m}{8}+1} f^3
\sqrt{H_{0}^2 \Gamma^{2 q+2}} \bigg(4 \sqrt[4]{6} \alpha +3 m \beta
\bigg) \bigg(\frac{H_{0}^2 \Gamma^{2 q+2}}{m^2}\bigg)^{\frac{3
m}{8}} + \big(\sqrt{H_{0}^2 \Gamma^{2 q+2}}
\\\nonumber
&\times \big(4 \sqrt[4]{6} \alpha +3 m \beta \big)-12 d\big)
\bigg(\frac{H_{0}^2 \Gamma^{2 q+2}}{m^2}\bigg)^{\frac{3 m}{8}}
+\bigg[2^{\frac{3 m}{8}-3} 3^{\frac{3 m}{8}+1} H_{0}^2 \Gamma^{2
q+2}  \sqrt{H_{0}^2 \Gamma^{2 q+2}}
\\\nonumber
&\times \big(4 \sqrt[4]{6} \alpha +3 m \beta \big) -12 d
\bigg(\frac{H_{0}^2 \Gamma^{2 q+2}}{m^2}\bigg)^{\frac{3
m}{8}-1}\bigg]\bigg[m\bigg]^{-1}\bigg\} \bigg(2^{\frac{3 m}{8}+1}
3^{\frac{3 m}{8}+2} a \sqrt{d} \big(3 H_{0}
\\\nonumber
&\times m (3 m+8) \Gamma^{q+1} -48 H_{0}^2 \bigg(12 H_{0}^2
\Gamma^{2 q+2}-3 H_{0} m \Gamma^{2 q+2}\bigg) \Gamma^{2 q+2} -432
H_{0}^4 m \Gamma^{4 q+4}\big)
\\\nonumber
&-6^{\frac{3 m}{8} +\frac{1}{2}} (3 m+4)\sqrt{H_{0}^2 \Gamma^{2
q+2}} \bigg(\frac{H_{0}^2 \Gamma^{2 q+2}}{m^2}\bigg)^{\frac{3 m}{8}}
\bigg(324 \sqrt{6} b \bigg(H_{0}^2 \Gamma^{2 q+2}\bigg)^{3/2}+108
\sqrt{6}
\\\nonumber
&\times \big(H_{0}^2 \Gamma^{2 q+2}\big)^{3/2} -72 6^{3/4} H_{0}^3
\Gamma^{3 q+3} \alpha-1728\ 6^{3/4} H_{0}^6 \Gamma^{6 q+6} \alpha
+864\ 6^{3/4} H_{0}^5 \Gamma^{6 q+6}
\\\nonumber
&\times \alpha -54 \sqrt{6} H_{0}^3 m \Gamma^{3 q+3} \beta -\sqrt{6}
H_{0}^6 m \Gamma^{6 q+6} \beta +\sqrt{6} H_{0}^5 m \Gamma^{6 q+6}
\beta +H_{0} \Gamma^{q+1}
\\\nonumber
&+H_{0}^3 \Gamma^{4 q+4} c_1\big)\big)\bigg] \bigg[18 H_{0}^2 (3
m+4) \big(6^{\frac{3 m}{8}+1} H_{0}^2 \Gamma^{2 q+2} \big\{6 f^3
\sqrt{H_{0}^2 \Gamma^{2 q+2}} \bigg(4 \sqrt[4]{6} \alpha +\beta
\bigg)
\\\nonumber
&\times\bigg(\frac{H_{0}^2 \Gamma^{2 q+2}}{m^2}\bigg)^{\frac{3
m}{8}} +2^{\frac{3 m}{8}}3^{\frac{3 m}{8}+2} a \sqrt{d} f^3
m\bigg)^2\bigg] -\bigg[f^3 \Gamma^{-2 q-2} \bigg(-6^{\frac{3
m}{8}+\frac{1}{2}} (3 m+4)
\\\nonumber
&\times \sqrt{H_{0}^2 \Gamma^{2 q+2}} \bigg(-12 6^{3/4} H_{0} \alpha
\Gamma^{q+1} - \sqrt{6} H_{0} m \beta \Gamma^{q+1}+ H_{0}^2 c_1
\Gamma^{2 q+2}+ H_{0} c_1 \Gamma^{2 q+2}
\\\nonumber
&-\ 6^{3/4} H_{0}^4 \alpha \Gamma^{4 q+4}+\ 6^{3/4} \times H_{0}^3
\alpha  \Gamma^{4 q+4}-\sqrt{6} H_{0}^4 m \beta \Gamma^{4 q+4}
+\sqrt{6} H_{0}^3 m \beta  \Gamma^{4 q+4}
\\\nonumber
&+ \sqrt{6} b \sqrt{H_{0}^2 \Gamma^{2 q+2}}+ \sqrt{6}
\sqrt{\text{ad}} \sqrt{H_{0}^2 \Gamma^{2 q+2}}\bigg)
\bigg(\frac{H_{0}^2 \Gamma^{2 q+2}}{m^2}\bigg)^{\frac{3 m}{8}}
\bigg[\frac{3 m}{8}\ +2^{\frac{3 m}{8}+1} 3^{\frac{3 m}{8}+2}
\\\nonumber
& \times\bigg(-96 H_{0}^2 \Gamma^{2 q+2}-72 H_{0}^2 m \Gamma^{2 q+2}
-8 \bigg(12 H_{0}^2 \Gamma^{2 q+2}-3 H_{0} m \Gamma^{2
q+2}\bigg)\bigg) -(3 m+4)
\\\nonumber
&\times \bigg(\frac{H_{0}^2 \Gamma^{2 q+2}}{m^2}\bigg)^{\frac{3
m}{8}} b \big(H_{0}^2 \Gamma^{2 q+2}\big)^{3/2} \bigg( \sqrt{6}+
\sqrt{6} \sqrt{\text{ad}} \bigg(H_{0}^2 \Gamma^{2 q+2}\bigg)^{3/2}
-72\ 6^{3/4} H_{0}^3
\\\nonumber
&\times\Gamma^{3 q+3} \alpha -6^{3/4} H_{0}^6 \Gamma^{6 q+6} \alpha
+\bigg(324 \sqrt{6} \bigg(324 \sqrt{6} 6^{3/4} H_{0}^5 \Gamma^{6
q+6} \alpha -54 \sqrt{6} H_{0}^3 m
\\\nonumber
&\times\Gamma^{3 q+3} \beta -\sqrt{6} H_{0}^6 m \Gamma^{6 q+6} \beta
+\sqrt{6} H_{0}^5 m \Gamma^{6 q+6} \beta +18 H_{0} \Gamma^{q+1}
c_1+H_{0}^3 \Gamma^{4 q+4} c_1\bigg)\bigg]
\\\nonumber
&\times\bigg[\sqrt{H_{0}^2 \Gamma^{2 q+2}}\bigg] -\bigg[2^{\frac{3
m}{8} -\frac{7}{2}} 3^{\frac{3 m}{8}+\frac{1}{2}} \sqrt{H_{0}^2
\Gamma^{2 q+2}} \bigg(\frac{H_{0}^2 \Gamma^{2
q+2}}{m^2}\bigg)^{\frac{3 m}{8}-1} \bigg(H_{0}^2 \Gamma^{2
q+2}\bigg)^{3/2}
\\\nonumber
&+\sqrt{6} \sqrt{\text{ad}} \bigg(H_{0}^2 \Gamma^{2 q+2}\bigg)^{3/2}
-\ 6^{3/4} H_{0}^3 \Gamma^{3 q+3} \alpha -1728\ 6^{3/4} H_{0}^6
\Gamma^{6 q+6} \alpha +864\ 6^{3/4}
 \\\nonumber
&\times H_{0}^5 \Gamma^{6 q+6} \alpha -54 \sqrt{6} H_{0}^3 m
\Gamma^{3 q+3} \beta -1296 \sqrt{6} H_{0}^6 m \Gamma^{6 q+6} \beta
+648 \sqrt{6} H_{0}^5 m \Gamma^{6 q+6} \beta
\\\nonumber
&+18 H_{0} \Gamma^{q+1} c_1+216 H_{0}^3 \Gamma^{4 q+4}
c_1\bigg)\bigg]\bigg[m\bigg]^{-1}\bigg] \bigg[18 H_{0}^2 (3 m+4)
\bigg(6^{\frac{3 m}{8}+1} H_{0}^2 \Gamma^{2 q+2} \bigg(6 f^3
\\\nonumber
&\times \sqrt{H_{0}^2 \Gamma^{2 q+2}} \bigg(4 \sqrt[4]{6} \alpha +3
m \beta \bigg)-12 d\bigg ) \bigg(\frac{H_{0}^2 \Gamma^{2
q+2}}{m^2}\bigg)^{\frac{3 m}{8}} +2^{\frac{3 m}{8}} 3^{\frac{3
m}{8}+2} a \sqrt{d} f^3 m\bigg)\bigg]^{-1}
\\\nonumber
& -3 \Gamma^{-q-1} \bigg(f^3 \Gamma^{-2 q-2} \bigg(\bigg[2^{\frac{3
m}{8}-2} 3^{\frac{3 m}{8}+1}  \alpha +3 m \beta \bigg)
\bigg(\frac{H_{0}^2 \Gamma^{2 q+2}}{m^2}\bigg)^{\frac{3
m}{8}}\bigg]\bigg[\sqrt{H_{0}^2 \Gamma^{2 q+2}}\bigg]^{-1}
\\\nonumber
&+\bigg[2^{\frac{3 m}{8}-4} 3^{\frac{3 m}{8}} \bigg(\frac{3
m}{8}-1\bigg) H_{0}^2 \Gamma^{2 q+2} \bigg(6 f^3 \sqrt{H_{0}^2
\Gamma^{2 q+2}} \big(4 \sqrt[4]{6} \alpha +3 m \beta \big)-\bigg
(\frac{H_{0}^2 }{m^2}\bigg)^{\frac{3 m}{8}-2}\bigg]
\\\nonumber
&\times\bigg[m^3\bigg]^{-1} +2^{\frac{3 m}{8}-3} 3^{\frac{3 m}{8}+1}
f^3 \sqrt{H_{0}^2 \Gamma^{2 q+2}} \big(4 \sqrt[4]{6} \alpha +3 m
\beta \big) \bigg (\frac{H_{0}^2 \Gamma^{2 q+2}}{m^2}\bigg)^{\frac{3
m}{8}-1} +\bigg[2^{\frac{3 m}{8}-3}
\\\nonumber
&\times 3^{\frac{3 m}{8}} \bigg(6 f^3 \sqrt{H_{0}^2 \Gamma^{2 q+2}}
(4 \sqrt[4]{6} \alpha +3 m \beta \big) -12 d\bigg)
\bigg(\frac{H_{0}^2 \Gamma^{2 q+2}}{m^2}\bigg)^{\frac{3
m}{8}-1}\bigg]\bigg[m\bigg]^{-1}\big(2^{\frac{3 m}{8}+1}
\\\nonumber
&\times 3^{\frac{3 m}{8}+2} a \sqrt{d} \bigg(3 H_{0} m
\Gamma^{q+1}-48 H_{0}^2 \bigg(12H_{0}^2 \Gamma^{2 q+2} -3 H_{0} m
\Gamma^{2 q+2}\bigg) \Gamma^{2 q+2} 432 H_{0}^4 m
\\\nonumber
&\times \Gamma^{4 q+4}\bigg)-6^{\frac{3 m}{8} +\frac{1}{2}} -(3 m+4)
\sqrt{H_{0}^2 \Gamma^{2 q+2}} \bigg(\frac{H_{0}^2 \Gamma^{2
q+2}}{m^2}\bigg)^{\frac{3 m}{8}} \bigg(324 \sqrt{6} b \bigg(H_{0}^2
\Gamma^{2 q+2}\bigg)^{3/2}
\\\nonumber
&+108 \sqrt{6} \sqrt{\text{ad}} \bigg(H_{0}^2 \Gamma^{2
q+2}\bigg)^{3/2}-72\ 6^{3/4} H_{0}^3 \Gamma^{3 q+3} \alpha -1728\
6^{3/4}H_{0}^6 \Gamma^{6 q+6} \alpha +864\ 6^{3/4}
\\\nonumber
&\times h^5 \Gamma^{6 q+6} \alpha -54 \sqrt{6} h^3 m \Gamma^{3 q+3}
\beta -1296 \sqrt{6} h^6 m \Gamma^{6 q+6} \beta +648 \sqrt{6} h^5
\times m \Gamma^{6 q+6} \beta
\\\nonumber
& +18 h \Gamma^{q+1} c_1 +216 h^3 \Gamma^{4 q+4} c_1\bigg)\bigg)18
h^2 (3 m+4) \bigg(6^{\frac{3 m}{8}+1} h^2 \Gamma^{2 q+2} \bigg(6 f^3
\sqrt{h^2 \Gamma^{2 q+2}}
\\\nonumber
&\times \big(4 \sqrt[4]{6} \alpha +3 m \beta \big) -12 d\bigg)
\bigg(\frac{h^2 \Gamma^{2 q+2}}{m^2}\bigg)^{\frac{3
m}{8}}+2^{\frac{3 m}{8}} 3^{\frac{3 m}{8}+2} a \sqrt{d} f^3 m\big)^2
+f^3 \Gamma^{-4 q-4}
\\\nonumber
&\times \big(-6^{3 m}{8} +\frac{1}{2} (3 m+4) \sqrt{h^2 \Gamma^{2
q+2}} \bigg(-6^{3/4} H_{0} \alpha \Gamma^{q+1}-9 \sqrt{6} H_{0} m
\beta \Gamma^{q+1} +108
\\\nonumber
&\times H_{0}^2 c_1 \Gamma^{2 q+2} + H_{0} c_1 \Gamma^{2 q+2} -\
6^{3/4} H_{0}^4 \alpha \Gamma^{4 q+4} +\ 6^{3/4} H_{0}^3 \alpha
\Gamma^{4 q+4} -\sqrt{6} H_{0}^4 m \beta  \Gamma^{4 q+4}
\\\nonumber
&+\sqrt{6} H_{0}^3 m \beta \Gamma^{4 q+4} +81 \sqrt{6} b
\sqrt{H_{0}^2 \Gamma^{2 q+2}}+27 \sqrt{6} \sqrt{\text{ad}}
\sqrt{H_{0}^2 \Gamma^{2 q+2}}\bigg ) \bigg(\frac{H_{0}^2 \Gamma^{2
q+2}}{m^2}\bigg)^{\frac{3 m}{8}}
\\\nonumber
&\times+2^{\frac{3 m}{8}+1} 3^{\frac{3 m}{8}+2} a \sqrt{d} \bigg(-96
H_{0}^2 \Gamma^{2 q+2}-72H_{0}^2 m \Gamma^{2 q+2} -8 \bigg(12
H_{0}^2 \Gamma^{2 q+2} -3 H_{0} m
\\\nonumber
&\times\Gamma^{2q+2}-2^{\frac{3 m}{8}-\frac{3}{2}} 3^{\frac{3
m}{8}-\frac{1}{2}} (3 m+4) \bigg(\frac{H_{0}^2 \Gamma^{2
q+2}}{m^2}\bigg)^{\frac{3 m}{8}} \bigg(324 \sqrt{6} b \bigg(H_{0}^2
\Gamma^{2 q+2}\bigg)^{3/2} +108
\\\nonumber
&\times \sqrt{6} \sqrt{\text{ad}} \bigg(H_{0}^2 \Gamma^{2
q+2}\bigg)^{3/2} -72\ 6^{3/4} H_{0}^3 \Gamma^{3 q+3} \alpha -
6^{3/4} H_{0}^6 \Gamma^{6 q+6} \alpha +6^{3/4} H_{0}^5
\\\nonumber
&\times\Gamma^{6 q+6} \alpha -54 \sqrt{6} H_{0}^3 m \Gamma^{3 q+3}
\beta -1296 \sqrt{6} H_{0}^6 m \Gamma^{6 q+6} \beta +648 \sqrt{6}
H_{0}^5 m \Gamma^{6 q+6} \beta
\\\nonumber
& +18 H_{0} \Gamma^{q+1} c_1\bigg].
\\\nonumber
s&=\frac{1}{2}+\bigg[\frac{H_{0} \alpha \Gamma^{q+1}}{72
\big(H_{0}^2 \Gamma^{2 q+2}\big)^{7/2}}-\big[5 H_{0} \big(3 \sqrt{6}
\log \big(6 H_{0}^2 \Gamma^{2 q+2}\big) \alpha -2 \sqrt{6} \alpha
+36 c_1\big)\Gamma^{q+1}\big]
\\\nonumber
&\times \big[\sqrt{6}\big(H_{0}^2 \Gamma^{2
q+2}\big)^{7/2}\big]^{-1} +\bigg[ \big(3 H_{0}^2 \Gamma^{2 q+2}
-H_{0} \Gamma^{2 q+2}\big)\big(\frac{\alpha \Gamma^{-2
q-2}}{\sqrt{6} H_{0}^2} +\frac{2 \sqrt{\frac{2}{3}} \beta
}{\sqrt{H_{0}^2 \Gamma^{2 q+2}}}\big)\bigg]
\\\nonumber
&\times\bigg\{\sqrt{H_{0}^2 \Gamma^{2 q+2}}\bigg\}^{-1}-8 \beta
-\frac{2 \alpha }{\sqrt{H_{0}^2 \Gamma^{2 q+2}}} -\frac{2 \sqrt{6}
c_1}{\sqrt{H_{0}^2 \Gamma^{2 q+2}}}-\frac{\alpha \log \big(6 H_{0}^2
\Gamma^{2q+2}\big)}{\sqrt{H_{0}^2 \Gamma^{2 q+2}}}
\\\nonumber
& -\bigg[\big(3 H_{0}^2 \Gamma^{2 q+2} -H_{0} \Gamma^{2 q+2}\big)
\bigg(\sqrt{6} \log \big(6 H_{0}^2 \Gamma^{2 q+2}\big) \alpha +
\alpha +\sqrt{H_{0}^2 \Gamma^{2 q+2}} \beta +12 c_1\bigg)\bigg]
\\\nonumber
&\times\bigg[\big(H_{0}^2 \Gamma^{2 q+2}\big)^{3/2}\bigg]^{-1}\bigg]
\bigg(-H_{0}^2 \beta \Gamma^{2 q+2}+(d+1)+b
\bigg(d+\frac{3}{2}\bigg) - \sqrt{H_{0}^2 \Gamma^{2 q+2}}\alpha
\\\nonumber
&\times \frac{1}{2}+a\sqrt{d}+\frac{d}{f^3}\bigg)\bigg]^{-1}
-\bigg[\bigg(-\frac{\alpha }{24 \sqrt{H_{0}^2 \Gamma^{2 q+2}}}
-\frac{\beta }{6}\bigg) \bigg(\bigg[H_{0} \bigg(3 \sqrt{6} \log \big
(6 H_{0}^2 \Gamma^{2 q+2}\big) \alpha
\\\nonumber
& -2 \sqrt{6} \alpha +36 c_1\bigg) \Gamma^{q+1}\bigg] \bigg[36
\sqrt{6} \bigg(H_{0}^2 \Gamma^{2 q+2}\bigg)^{5/2}\bigg]-48 H_{0}^2
\beta  \Gamma^{2 q+2}-\sqrt{H_{0}^2 \Gamma^{2 q+2}} c_1
\\\nonumber
& -12 \sqrt{H_{0}^2 \Gamma^{2 q+2}} \alpha +8 \sqrt{6} \sqrt{H_{0}^2
\Gamma^{2 q+2}} \beta +12 c_1\big)\bigg]\bigg[\sqrt{H_{0}^2
\Gamma^{2 q+2}}\bigg]^{-1}-48 a \sqrt{d}\big)\bigg]
\\\nonumber
&\times \bigg[-H_{0}^2 \beta \Gamma^{2 q+2} +\sqrt{\text{ad}} (d+1)
+b \bigg(d+\frac{3}{2}\bigg)-\frac{1}{2} \sqrt{H_{0}^2 \Gamma^{2
q+2}} \alpha+a\sqrt{d}+\frac{d}{f^3}\big)^2\bigg]
\\\nonumber
&+36 c_1\big(H_{0}^2 \Gamma^{2 q+2}\big)^{9/2}-\frac{H_{0} \alpha
\Gamma^{q+1}}{72 \big(H_{0}^2 \Gamma^{2 q+2}\big)^{9/2}}+
\sqrt{\frac{2}{3}} \big(3 H_{0}^2 \Gamma^{2 q+2} -H_{0} \Gamma^{2
q+2}\big)
\\\nonumber
&-\big(\alpha  \Gamma^{-4 q-4}\big)\big(6 \sqrt{6} H_{0}^4\big)^{-1}
{q+2}\bigg)^{3/2} +\bigg[\alpha  \log \big(6 H_{0}^2
\Gamma^{2q+2}\big)\bigg]\bigg[12 \big(H_{0}^2 \Gamma^{2 q+2}\big
)^{3/2}\bigg]^{-1}
\\\nonumber
&+\big(3 H_{0}^2 \Gamma^{2 q+2} -H_{0} \Gamma^{2 q+2}\big)
\big(\sqrt{6} \log \bigg(6H_{0}^2 \Gamma^{2 q+2}]\big ) \alpha +2
\sqrt{6} \alpha + \sqrt{h^2 \Gamma^{2 q+2}} \beta + 12 c_1
\\\nonumber
&\times{12 \sqrt{6} \big(h^2 \Gamma^{2 q+2}\big)^{5/2}} - \big(3 h^2
\Gamma^{2 q+2}-h \Gamma^{2 q+2}\big) \bigg(\frac{\alpha \Gamma^{-2
q-2}}{\sqrt{6} h^2} +\frac{2 \sqrt{\frac{2}{3}} \beta }{\sqrt{h^2
\Gamma^{2 q+2}}}\bigg){3 \big(h^2 \Gamma^{2 q+2}\big)^{3/2}}
\\\nonumber
&\times32 \bigg(-h^2 \beta \Gamma^{2 q+2} +\sqrt{\text{ad}} (d+1)+b
\bigg(d+\frac{3}{2}\bigg)-\frac{1}{2} \sqrt{h^2 \Gamma^{2 q+2}}
\alpha +a \sqrt{d}+\frac{d}{f^3}\bigg )
\\\nonumber
&\times-5 \bigg(-\frac{\alpha }{24 \sqrt{h^2 \Gamma^{2 q+2}}}
-\frac{\beta }{6}\bigg ) \bigg(\frac{h \alpha  \Gamma^{q+1}}{72
\big(h^2 \Gamma^{2 q+2}\big)^{7/2}} -\big(5 H_{0} \big(3 \sqrt{6}
\log \big(6 H_{0}^2 \Gamma^{2 q+2}\big)
\\\nonumber
&\times\alpha -2 \sqrt{6} \alpha +36 c_1\big)
\Gamma^{q+1}\bigg)\bigg(432 \sqrt{6} \big(H_{0}^2 \Gamma^{2
q+2}\big)^{7/2}\bigg)^{-1} +2 \sqrt{\frac{2}{3}} \bigg(3 H_{0}^2
\Gamma^{2 q+2}
\\\nonumber
&-\bigg[\alpha \log \big(6 H_{0}^2 \Gamma^{2
q+2}\big)\bigg]\bigg[\sqrt{H_{0}^2 \Gamma^{2 q+2}}\bigg]
-\bigg[\big(3 H_{0}^2 \Gamma^{2 q+2} -H_{0} \Gamma^{2 q+2}\big)
\big(6 H_{0}^2\Gamma^{2 q+2}\big) \alpha
\\\label{65}
&+2 \sqrt{6} \alpha +8 \sqrt{6} \sqrt{H_{0}^2 \Gamma^{2 q+2}} \beta
+12 c_1\bigg]\bigg[3 \sqrt{6} \bigg(H_{0}^2 \Gamma^{2
q+2}\bigg)^{3/2}\bigg]\bigg].
\end{align}

\section*{Appendix C: Determination of $\nu_{s}^{2}$ }
\renewcommand{\theequation}{C\arabic{equation}}
\setcounter{equation}{0}

\begin{align}\nonumber
\nu_{s}^{2}&=\bigg[\bigg(\frac{h^2 \Gamma ^{2
q+2}}{m^2}\bigg)^{-\frac{3 m}{8}}36 a \sqrt{d}-18 \bigg(\frac{h^2
\Gamma^{2 q+2}}{m^2}\bigg)^{\frac{3 m}{8}}2 \big(4 \sqrt[4]{6}
\alpha +3 m \beta \big) \big(h^2 \Gamma^{2 q+2}\big)^{3/2}
\\\nonumber
&+b (d-2)+\sqrt{\text{ad}} (d-1)\bigg[f^3 \Gamma^{-4 q-4} 2^{\frac{3
m}{8}+1} 3^{\frac{3 m}{8}+2} a \sqrt{d} \big(3 h m (3 m+8)
\Gamma^{q+1}
\\\nonumber
&-48 h^2 \big(12 h^2 \Gamma^{2 q+2}-3 h m \Gamma^{2 q+2}\big)
\Gamma^{2 q+2}-432 h^4 m \Gamma^{4 q+4}\big)-6^{\frac{3 m}{8}
+\frac{1}{2}} \sqrt{h^2 \Gamma^{2 q+2}}
\\\nonumber
&\times \bigg(\frac{H_{0}^2 \Gamma^{2 q+2}}{m^2}\bigg)^{\frac{3
m}{8}} 324 \sqrt{6} b \big(H_{0}^2 \Gamma^{2 q+2}\big)^{3/2}+108
\sqrt{6} \sqrt{\text{ad}} \big(H_{0}^2 \Gamma^{2 q+2}\big)^{3/2}
\\\nonumber
&-72\ 6^{3/4} H_{0}^3 \Gamma^{3 q+3} \alpha -1728\ 6^{3/4} H_{0}^6
\Gamma^{6 q+6} \alpha +864\ 6^{3/4} H_{0}^5 \Gamma^{6 q+6} \alpha
-54 \sqrt{6} H_{0}^3
\\\nonumber
&\times m \Gamma^{3 q+3} \beta -1296 \sqrt{6} H_{0}^6 m \Gamma^{6
q+6} \beta +648 \sqrt{6} H_{0}^5 m \Gamma^{6 q+6} \beta +18 H_{0}
\Gamma^{q+1} c_1
\\\nonumber
&+216 H_{0}^3 \Gamma^{4 q+4} c_1\bigg] \bigg[108 H_{0}^4 (3 m+4)
\big(6^{\frac{3 m}{8}+1} H_{0}^2 \Gamma^{2 q+2} \big(6 f^3
\sqrt{H_{0}^2 \Gamma^{2 q+2}} \big(4 \sqrt[4]{6} \alpha
\\\nonumber
&+3 m \beta \big) -12 d\big) \bigg(\frac{H_{0}^2 \Gamma^{2
q+2}}{m^2}\bigg)^{\frac{3 m}{8}} +2^{\frac{3 m}{8}} 3^{\frac{3
m}{8}+2} a \sqrt{d} f^3 m\big)\bigg]^{-1}+\bigg[f^3 \Gamma^{-2 q-2}
\\\nonumber
&\times2^{\frac{3 m}{8}} 3^{\frac{3 m}{8}+1} f^3 \sqrt{H_{0}^2
\Gamma^{2 q+2}} \big(4 \sqrt[4]{6} \alpha +3 m \beta \big)
\bigg(\frac{H_{0}^2 \Gamma^{2 q+2}}{m^2}\bigg)^{\frac{3
m}{8}}+\big(\sqrt{H_{0}^2 \Gamma^{2 q+2}}
\\\nonumber
&\times \big(4 \sqrt[4]{6} \alpha +3 m \beta \big) -12 d\big)
\bigg(\frac{H_{0}^2 \Gamma^{2 q+2}}{m^2}\bigg)^{\frac{3 m}{8}}
+\bigg[2^{\frac{3 m}{8}-3} 3^{\frac{3 m}{8}+1} H_{0}^2 (z+1) ^{2
q+2} \big(6 f^3
\\\nonumber
&\times\sqrt{H_{0}^2 \Gamma^{2 q+2}} \big(4 \sqrt[4]{6} \alpha-432
H_{0}^4 m \Gamma^{4 q+4}\big) +3 m \beta \big)\big(\frac{H_{0}^2
(z+1) ^{2 q+2}}{m^2}\big)^{\frac{3 m}{8}-1}\bigg]\bigg[m\bigg]^{-1}
\\\nonumber
& \times2^{\frac{3 m}{8}+1} 3^{\frac{3 m}{8}+2} a \sqrt{d}3 H_{0} m
(3 m+8) (z+1) ^{q+1}-48 H_{0}^2 \big(12 H_{0}^2 \Gamma^{2 q+2}-3
H_{0} m
\\\nonumber
&\times(z+1) ^{2 q+2}\big) \Gamma^{2 q+2}-6^{\frac{3 m}{8}
+\frac{1}{2}} (3 m+4) \sqrt{H_{0}^2 (z+1) ^{2 q+2}}
\bigg(\frac{H_{0}^2 \Gamma^{2 q+2}}{m^2}\bigg)^{\frac{3 m}{8}}
\\\nonumber
&- \big(H_{0}^2 (z+1) ^{2 q+2}\big)^{3/2} +108 \sqrt{6}
\sqrt{\text{ad}} \bigg (H_{0}^2 \Gamma^{2 q+2}\bigg)^{3/2}-72\
6^{3/4} H_{0}^3 (z+1)^{3 q+3} \alpha
\\\nonumber
&-1728\ 6^{3/4} H_{0}^6 \Gamma^{6 q+6} \alpha +864\ 6^{3/4} -54
\sqrt{6} H_{0}^3 m (z+1) ^{3 q+3} \beta -1296 \sqrt{6} H_{0}^6
\\\nonumber
&\times  m (z+1) ^{6 q+6} \beta +648 \sqrt{6} H_{0}^5 m \Gamma^{6
q+6} \beta +H_{0} \Gamma^{q+1} c_1 +H_{0}^3 (z+1) ^{4 q+4} c_1\bigg]
\\\nonumber
&\times\bigg[18 H_{0}^2 (3 m+4) \big(6^{\frac{3 m}{8}+1} H_{0}^2
\Gamma^{2 q+2} \big(6 f^3 \sqrt{H_{0}^2 \Gamma^{2 q+2}} \big (4
\sqrt[4]{6} \alpha +3 m \beta \big)-12 d\big)
\\\nonumber
&\times \bigg(\frac{H_{0}^2 \Gamma^{2 q+2}}{m^2}\bigg)^{\frac{3
m}{8}}+2^{\frac{3 m}{8}} 3^{\frac{3 m}{8}+2} a \sqrt{d} f^3
m\big)^2\bigg]^{-1} -\bigg[f^3 (z+1) ^{-2 q-2} \bigg(-6^{\frac{3
m}{8} +\frac{1}{2}}
\\\nonumber
& \times(3 m+4) \sqrt{H_{0}^2 \Gamma^{2 q+2}}-12 6^{3/4} H_{0}
\alpha  (z+1) ^{q+1}-9 \sqrt{6} H_{0} m \beta \Gamma^{q+1} +108
H_{0}^2
\\\nonumber
&\times c_1 (z+1) ^{2 q+2} +36 H_{0} c_1 \Gamma^{2 q+2}-432\ 6^{3/4}
H_{0}^4 \alpha (z+1) ^{4 q+4} +6^{3/4} H_{0}^3 \alpha \Gamma^{4 q+4}
\\\nonumber
&-324 \sqrt{6} H_{0}^4 m \beta (z+1) ^{4 q+4}+216 \sqrt{6} H_{0}^3 m
\beta \Gamma^{4 q+4}\bigg]+b \sqrt{H_{0}^2 (z+1) ^{2 q+2}}
\\\nonumber
&+27 \sqrt{6} \sqrt{\text{ad}} \sqrt{H_{0}^2 \Gamma^{2 q+2}}
\bigg(\frac{H_{0}^2 (z+1) ^{2 q+2}}{m^2}\bigg)^{\frac{3 m}{8}}
+2^{\frac{3 m}{8}+1} 3^{\frac{3 m}{8}+2} a \sqrt{d} \bigg [-96
H_{0}^2
\\\label{66}
&\times \Gamma^{2 q+2}-72 H_{0}^2 m (z+1) ^{2 q+2} -8 \big(12
H_{0}^2 (z+1) ^{2 q+2}-3 H_{0} m \Gamma^{2 q+2}\big)\bigg].
\end{align}
\\
\textbf{Data availability:} No new data were generated or analyzed
in support of this research.

\end{document}